\documentclass[twocolumn,aps]{revtex4-1}
\usepackage{amsmath}
\usepackage{amssymb}
\usepackage{widetable}
\usepackage{graphics}
\usepackage[T1]{fontenc}
\usepackage{epstopdf}
\usepackage{dcolumn}
\usepackage{bm}
\usepackage{rotating}
\usepackage{bbm}
\usepackage{verbatim}
\usepackage{txfonts}
\usepackage[cspex,bbgreekl]{mathbbol}
\usepackage{subfigure}
\usepackage{tabularx}
\usepackage{graphicx}
\usepackage{hyperref}
\hypersetup{
  colorlinks=true,
  linkcolor=blue,
  citecolor=red,
  linktoc=all,
  pdftitle=Dielectric matrix formulation of correlation energies in the Random Phase Approximation (RPA): inclusion of exchange effects,
  pdfdisplaydoctitle=true,
  pdfpagelayout=SinglePage,
  pdfstartview=Fit,
  pdfstartpage=1,
  bookmarksopen=false
}
\newcommand{\T}{\ensuremath{\text{T}}}
\newcommand{\Det}{\text{Det}}
\newcommand{\Log}{\text{Log}}
\newcommand{\Tr}{\text{Tr}}
\newcommand{\tr}{\text{tr}}
\newcommand{\mhalf}{{-}\frac{1}{2}}
\newcommand{\half}{\frac{1}{2}}
\renewcommand{\i}{\mathrm{i}}
\newcommand{\bra}[1]{\ensuremath{\langle #1 \vert}}

\newcommand\bb[1] {\ensuremath{\mathbb{#1}}}
\renewcommand\b[1]{\ensuremath{\mathbf{#1}}}

\newcommand\B[1] {\ensuremath{\pmb #1}}
\begin{document}
\author{Bastien Mussard}
\email{bastien.mussard@upmc.fr}
\affiliation{
Sorbonne Universit\'es, UPMC Univ Paris 06, CNRS, Laboratoire de Chimie Th\'eorique, F-75005 Paris, France}
\affiliation{
Sorbonne Universit\'es, UPMC Univ Paris 06, Institut du Calcul et de la Simulation, F-75005, Paris, France}

\author{Dario Rocca}
\affiliation{
Universit\'e de Lorraine, CRM\textsuperscript{2}, UMR 7036, Vandoeuvre-l\`es-Nancy, F-54506, France}
\affiliation{
CNRS, CRM\textsuperscript{2}, UMR 7036, Vandoeuvre-l\`es-Nancy, F-54506, France}

\author{Georg Jansen}
\affiliation{
Fakult\"at f\"ur Chemie, Universit\"at Duisburg-Essen, D-45117 Essen, Germany}

\author{J\'anos G.~\'Angy\'an}
\affiliation{
CNRS, CRM\textsuperscript{2}, UMR 7036, Vandoeuvre-l\`es-Nancy, F-54506, France}
\affiliation{
Universit\'e de Lorraine, CRM\textsuperscript{2}, UMR 7036, Vandoeuvre-l\`es-Nancy, F-54506, France}
\title[RPA: Dielectric matrix formulation]{Dielectric matrix formulation of correlation energies in the Random Phase Approximation (RPA): inclusion of exchange effects}
\begin{abstract}
Starting from the general expression for the ground state correlation energy in the adiabatic connection fluctuation dissipation theorem (ACFDT) framework, it is shown that the dielectric matrix formulation, which is usually applied to calculate the direct random phase approximation (dRPA) correlation energy, can be used for alternative RPA expressions including exchange effects. Within this famework, the ACFDT analog of the second order screened exchange (SOSEX) approximation leads to
a logarithmic formula for the correlation energy similar to the direct RPA expression. Alternatively, the contribution of the exchange can be included in the
kernel used to evaluate the response functions. In this case the use of an approximate kernel is crucial to simplify the formalism and to obtain a
correlation energy in logarithmic form. Technical details of the implementation of these methods are discussed and it is shown that one can take advantage of
density fitting or Cholesky decomposition techniques to improve the computational efficiency; a discussion on the numerical quadrature made on the frequency variable is also provided. A series of test calculations on atomic correlation
energies and molecular reaction energies shows that exchange effects are instrumental to improve over
direct RPA results.
\end{abstract}

\maketitle
\section{Introduction}\label{sec:introduction}

In the past few years, random phase approximation (RPA) approaches, which belong to the methods on the 5th rung of Jacob's ladder~\cite{Perdew:01f}, have been on the way of becoming a practical tool to construct correlation energy functionals~\cite{Furche:01b,Fuchs:02,Miyake:02,Furche:05,Fuchs:05,Furche:08,Harl:08,Lu:09,LiYan:10,Gruneis:09,Nguyen:09,Nguyen:10,Paier:10,Hesselmann:10,Hesselmann:11x,Eshuis:12b,Hesselmann:12,Ren:12,Toulouse:09,Janesko:09,Janesko:09a,rocca:14}.
Specifically, within the framework of the adiabatic-connection fluctuation-dissipation-theorem (ACFDT), the electronic correlation energy is
expressed in terms of dynamical (frequency-dependent) linear response functions and the electron-electron interaction is gradually switched on from  the independent particle reference to the fully interacting state using an adiabatic connection parameter. Therefore the correlation energy takes the form of a multiple integral involving both the frequency and the adiabatic connection parameters. Instead of the exact linear response function, convenient approximations, like the RPA, are used.

Among the numerous possible ways  to express the RPA ground state correlation energy, depending on the order in which the analytical and/or numerical frequency- and interaction strength-integrations are performed, one may cite (1) the \textit{density matrix formulation}~\cite{Furche:01b}, which leads to expressions involving numerical integration over the interaction strength and (2) the \textit{dielectric matrix formulation} which involves a numerical integral over frequency of a logarithmic expression involving the dynamical dielectric function. This approach has mainly been used by solid state theorists~\cite{Miyake:02,Harl:09,Lu:09,Nguyen:09} and recently adapted also in a density fitting framework by Furche \textit{et al.} \cite{Eshuis:10}. Within this formulation the direct calculation of the lowest eigenvalues and eigenvectors of the dielectric matrix~\cite{Wilson:08} can be used to achieve a more compact representation of dielectric functions~\cite{Lu:09,Nguyen:09,Nguyen:10,LiYan:10,rocca:14,kaoui2016random}. Finally, there are approaches which avoid numerical integration altogether, like (3) the \textit{plasmon formula}, obtained after a double analytical integration on both the frequency and the interaction strength~\cite{Furche:08}. An elegant way to obtain the plasmon expression consists in solving the algebraic Riccati-equations of the RPA problem~\cite{Sanderson:65,Szabo:77}. This method has been shown to be strictly equivalent to a coupled cluster doubles approach in the ring-approximation (rCCD)~\cite{Toulouse:11,Scuseria:08}. Modifications and approximations to the rCCD equations and energy expression has led to a whole class of additional RPA-based approaches for the ground state correlation energy~\cite{Szabo:77,Lotrich:11,Klopper:11,Hesselmann:12}.

The RPA problem  can be set up either with the inclusion of nonlocal exchange effects (leading to a class of approximations denoted RPAx) or by restricting the coupling between independent particle responses to the direct Hartree interaction (leading to the class denoted dRPA). Additionally, according to the classification and nomenclature proposed in Ref.~\onlinecite{Angyan:11}, RPA methods for the ground-state correlation energy can be sought in two main flavors: either by the complete neglect of the exchange integrals, \textit{i.e.}\ by taking the contraction of the dRPA response function with non-antisymmetrized two-electron integrals (dRPA-I), or by a full consideration of non-local exchange, when antisymmetrized two-electron integrals are contracted with the RPAx response function matrix elements (RPAx-II). This latter approach has been followed by some early works in quantum chemistry, based on Hartree-Fock orbitals~\cite{Oddershede:78,Szabo:77,Ostlund:71,McLachlan:64}, while dRPA is usually the method of choice in the density functional context, starting from Kohn-Sham orbitals. Partial inclusion/omission of nonlocal exchange leads to "mixed" methodologies, like dRPA-II and RPAx-I.

It has been suggested that certain shortcomings of the dRPA correlation energy can be remedied by including nonlocal exchange interactions in a perturbative way, \textit{i.e.}\ with the dRPA polarization propagator being contracted with a list of  fully antisymmetrized  two-electron integrals~\cite{Gruneis:09,Paier:10}. We can mention, in this aspect, the SOSEX (second order  screened exchange) corrections, which have been formulated
originally within the rCCD formalism. It has been pointed out that "it is difficult to motivate this approximation in the framework of ACFDT"~\cite{Gruneis:09}. This situation seems to be somewhat paradoxical, since the plasmon formula, identical to the rCCD correlation energy expression, can be derived from ACFDT in a straightforward manner~\cite{Furche:08}, and there is no fundamental reason to think that an analogous derivation is impossible for SOSEX. In fact, it has been
shown that a very similar perturbative screened exchange formula, which has been designated by the acronym dRPA-IIa, can be obtained within the density matrix formulation of RPA~\cite{Angyan:11}. Although this expression gives correlation energies numerically very close to the rCCD-based SOSEX, it has been proven that they are not strictly identical~\cite{Jansen:10}. Recently, in a similar but different fashion, the AXK introduced by Bates \textit{et al.} reduces the self-interaction
error and improves the description of static correlation over dRPA.

In this work we discuss different approximations to efficiently include exchange effects within the dielectric matrix formulation of the RPA correlation energy.
In a quantum chemical context these methodologies provide an alternative to the ring-CCD-based RPA formalism.
Additionally, these methods are of significant interest for the solid state physics community, where the
dielectric matrix approach is almost exclusively used.
As it will be shown in the following, the dRPA-IIa approximation can be derived also in the dielectric matrix formulation of RPA, leading to an alternative algorithm to calculate the SOSEX-like dRPA-IIa energy.
At the same time we have access to the conventional MP2 energy, which corresponds to the lowest non-vanishing term in 
the series expansion of the dRPA-IIa correlation energy.
We note that approximations in the RPAx class lead to other correlation energy estimates, which can be also adapted to the logarithmic formulation. Specifically, the RPAx-Ia approximation is derived, which includes only the particle-hole contribution to the exchange
kernel. It will be shown that this last approach provides the most promising results in the numerical tests considered in this work.

In Sec.~\ref{sec:dRPA}, the working equations for the well-known dielectric matrix formulation of the direct RPA method will be derived from the ACFDT.
This version is particularly simple because of the complete neglect of the exchange effects. Incorporation of exchange leads to a more complicated expression and
one is constrained to apply various additional approximations to obtain practical expressions.
As discussed in Sec.~\ref{RPASOSEX}, one possibility is a SOSEX-like correlation energy expression, which is the dielectric matrix
formulation of the dRPA-IIa variant.
In Sec.~\ref{sec:RPAx} we will discuss an additional variant, which includes exchange effects in the response function.
The possible practical implementations of these methodologies are discussed in some details in Sec.~\ref{computational}.
Several atomic and molecular systems will be used for numerical illustration of the formalism, in Sec.~\ref{numerical}.
Finally, Sec.~\ref{conclusions} contains our conclusions.

\section{RPA versions from ACFDT and their dielectric matrix formulation}
\label{sec:RPA}

In the adiabatic-connection fluctuation-dissipation theorem (ACFDT) approach the correlation energy reads as~\cite{Langreth:75,Langreth:77}:

\begin{align}\label{eq:ACFDTformula}
 E^{\text{ACFDT}}_{\text{c}}&=
 \mhalf
 \int _0^1 \!\! d\alpha
 \int_{-\infty}^{\infty}\!\! \frac{d\omega}{2\pi}\,
 \Tr
 \left\{ \bb{\Pi}_{\alpha } (\i\omega) \, {\bb{G}}-
         \bb{\Pi}_0 (\i\omega)\,{\bb{G}}
 \right\}
,\end{align}
where $\bb{G}$ is the matrix representation of the interaction which contains two-electron integrals (see Eqs.~(\ref{eq:Vinteraction}) and~(\ref{eq:Winteraction})).
In Eq.~\ref{eq:ACFDTformula} the four-index matrix representation of the dynamical polarization propagator $\bb{\Pi}_{\alpha } (\i\omega)$ 
(also called density matrix response function) at interaction strength $\alpha$ is obtained from a Dyson-like equation,

\begin{align}
\label{eq:Dysoninv}
   \bb{\Pi}_\alpha(\i\omega)=
   (\bb{I}-\alpha\,\bb{\Pi}_0(\i\omega)\,
    \bb{F})^{-1}\bb{\Pi}_0(\i\omega)
,\end{align}
where $\bb{F}$ is the interaction kernel matrix (see Eqs.~(\ref{eq:Vinteraction}) and~(\ref{eq:DysoninvRPAx})).
In the two previous equations, the polarization propagator of the non-interacting reference system, $\bb{\Pi}_0 (\i\omega)$, is given by:

\begin{align}\label{eq:Pi0def}
   \bb{\Pi}_0(\i\omega) =-
   (\bb{\Lambda}_0 - \i\omega\,\bb{\Delta})^{-1}
,\end{align}
with

\begin{align}
\label{eq:Lambda0_Delta}
\bb{\Lambda}_0=
\begin{pmatrix}
 \B{\epsilon}&   \b{0}  \\
\b{0}& \B{\epsilon}
\end{pmatrix}
\qquad\quad \text{and}\qquad\quad
\bb{\Delta}=
\begin{pmatrix}
 \b{1}&   \b{0}  \\
\b{0}& -\b{1}
\end{pmatrix}
.\end{align}
In the random phase approximation (RPA), the elements of the
$\bb{\Lambda}_0$ matrix are the independent one-particle
excitation energies:
$(\B{\epsilon})_{ia,jb}\!=\! \epsilon_{ia}\delta_{ij}\delta_{ab}$, where
$\epsilon_{ia}\!=\!\epsilon_{a}\!-\!\epsilon_i$,
$\epsilon_a$ is the energy of a virtual orbital and
$\epsilon_i$ the energy of an occupied orbital.
Here and in the following we assume that a finite-dimensional basis
set is used for the representation of occupied and virtual
orbitals.

The polarization propagator of the non-interacting
reference system~\cite{Oddershede:78} then reads:

\begin{align}
\label{eq:Pi0viaPi0pm}
\bb{\Pi}_0(\i\omega)=
\begin{pmatrix}
  \b{\Pi}_0^+(\i\omega) & \b{0} \\
  \b{0} & \b{\Pi}_0^-(\i\omega)
\end{pmatrix}
,\end{align}
which introduces compact notations
for the diagonal blocks, $\b{\Pi}_0^+(\i\omega)=-\left(\B{\epsilon}-\i\omega\b{1}\right)^{-1}$ and $\b{\Pi}_0^-(\i\omega)=-\left(\B{\epsilon}+\i\omega\b{1}\right)^{-1}$.

The
dimensions of the matrices appearing in  Eq.~(\ref{eq:ACFDTformula}) are $(2 N_\text{exc}\times 2 N_\text{exc})$, where $N_\text{exc}$
is the number of the products between occupied and virtual orbitals.
However, since we need only the trace for the correlation energy, the same result
can be obtained using matrices with a dimension reduced to $(N_\text{exc}\times N_\text{exc})$, as it will be shown below.

In several earlier works~\cite{Harl:Thesis,Hesselmann:10} the derivation of the logarithmic energy expression have been based on a Taylor expansion of matrix functions (namely the inverse and the logarithmic functions).
Such a procedure raises some problems of general validity, since the conditions for an absolute convergence of the series expansions cannot be always satisfied.
Below, we propose an alternative derivation using matrix functions, which requires only the weaker condition that the matrix be diagonalizable.
The general definition of a matrix function is

\begin{align}
\label{eq:fofm}
f({\b{A})=\b{Q}\,f(\b{D})\,\b{Q}^{-1}}
,\end{align}
where
$\b{A}=\b{Q}\b{D}\b{Q}^{-1}$ is the eigendecomposition of the matrix $\b{A}$.
Hence, all along the manuscript, the matrices $\b{D}$ are diagonal matrices of eigenvalues and $\b{Q}$ are eigenvectors.

\subsection{Direct RPA}\label{sec:dRPA}

In the direct RPA  (dRPA), the nonlocal Hartree-Fock exchange is not taken into account in the interaction kernel $\bb{F}$ of the Dyson-like equation in Eq.~(\ref{eq:Dysoninv}).
Furthermore, we will first consider an interaction matrix $\bb{G}$ constituted of non-antisymmetrized two-electron integrals in Eq.~(\ref{eq:ACFDTformula}).
This yields:

\begin{align}\label{eq:Vinteraction}
\bb{F}=\bb{G}={\bb{V}}=
\begin{pmatrix}
 \b{K}&   \b{K}  \\
\b{K}& \b{K}
\end{pmatrix}
,\end{align}
where $K_{ia,jb}\!\! = \!\!\bra{ij} ab\rangle$  are the non-antisymmetrized
two-electron integrals (physicist's notation) in spin-orbitals.
In this situation, one only has to deal with the matrix product $\bb{\Pi}_0(\i\omega)\bb{V}$ (see Eqs.~(\ref{eq:ACFDTformula}) and ~(\ref{eq:Dysoninv})).
Considering the block-structure of
$\bb{\Pi}_0{\bb{V}}$,

\begin{align}
\bb{\Pi}_0\bb{V}=
\begin{pmatrix}
  \b{\Pi}_0^+\b{K} & \b{\Pi}_0^+\b{K}\\
  \b{\Pi}_0^-\b{K} & \b{\Pi}_0^-\b{K}
\end{pmatrix}
,\end{align}
application of the following unitary transformation:

\begin{align}\label{eq:unitary}
\bb{U}=\frac{1}{\sqrt{2}}
\begin{pmatrix}
\b{I}&~~\b{I}\\
\b{I}&- \b{I}
\end{pmatrix}
,\end{align}
to the integrand of Eq.~(\ref{eq:ACFDTformula}) yields:

\begin{align}
&\Tr\{(1-\alpha\bb{\Pi}_0(\i\omega)\bb{V})^{-1}\bb{\Pi}_0\bb{V}-\bb{\Pi}_0\bb{V}\}
\nonumber\\&\quad
=
\tr\{(1-\alpha\b{\Pi}_0(\i\omega)\b{K})^{-1}\b{\Pi}_0\b{K}-\b{\Pi}_0\b{K}\}
,\end{align}
where we use
the notation $\b{{\Pi}}_0(\i\omega )=\b{{\Pi}}_0^+(\i\omega )+\b{{\Pi}}_0^-(\i\omega )$.
The notation $\tr\{\b{X}\}$
emphasizes that we have reduced the matrix dimensions as compared to $\Tr\{\bb{X}\}$
(the same applies later to $\det(\b{X})$/$\Det(\bb{X})$ and to $\log(\b{X})$/$\Log(\bb{X})$).
Defining the function
$\b{\Pi}_{\alpha } (\i\omega)$ that satisfies the $(N_\text{exc}\times N_\text{exc})$ dimension-reduced Dyson-like equation,

\begin{align}\label{eq:Dysoninvsmall}
  \b{\Pi}_{\alpha } (\i\omega) = \bigl(\b{I} -\alpha \,  \b{\Pi}_0 (\i\omega)\b{K}
  \bigr)^{-1} \b{\Pi}_0 (\i\omega)
,\end{align}
we retrieve
the analog of Eq.~(\ref{eq:ACFDTformula}) with dimension-reduced matrices

\begin{align}\label{eq:dRPAIsmall}
 E^{\text{dRPA-I}}_{\text{c}}&=
 \mhalf
 \int _0^1 \!\! d\alpha
 \int_{-\infty}^{\infty}\!\! \frac{d\omega}{2\pi}\,
 \tr
 \left\{ \b{\Pi}_{\alpha } (\i\omega) {\b{K}}-
         \b{\Pi}_0 (\i\omega)  {\b{K}}
 \right\}
.\end{align}
Equation~(\ref{eq:dRPAIsmall}) can be further transformed either by an
analytical frequency-integration,
which leads to  the density matrix expression of the dRPA correlation energy
(this has already been discussed in a previous publication~\cite{Angyan:11}, 
and is briefly recalled in Appendix~\ref{app:dRPADM}), 
or by an analytical integration over the adiabatic connection parameter.
This will be done by considering the eigenvalue decomposition of
$\b{{\Pi}}_0(\i\omega )\,\b{K}$. Using the dimension-reduced Dyson-like equation (Eq.~\ref{eq:Dysoninvsmall}),  
we express
$\b{\Pi}_\alpha(\i\omega){\b{K}}$ seen in Eq.~(\ref{eq:dRPAIsmall}) as a matrix function of $\b{\Pi}_0(\i\omega){\b{K}}$:

\begin{align}
\label{eq:PialphaK}
\b{\Pi}_\alpha(\i\omega){\b{K}} =
\b{Q}(\i\omega) \bigl(\b{I}-\alpha \b{D}(\i\omega)\bigr)^{-1} \b{D}(\i\omega) \b{Q}^{-1}(\i\omega)
,\end{align}
where, as stated before, $\b{D}(\i\omega)$ is the diagonal matrix of the eigenvalues of $\b{\Pi}_0(\i\omega){\b{K}}$ and $\b{Q}(\i\omega)$ contains its eigenvectors.
Using the cyclic invariance of the trace and denoting by $d_{ia}$ the $ia$-th diagonal element of $\b{D}$, Eq.~(\ref{eq:dRPAIsmall}) becomes

\begin{align}
 E^{\text{dRPA-I}}_{\text{c}}\!\!&=
 \mhalf
 \int_{-\infty}^{\infty}\!\! \frac{d\omega}{2\pi}\,
 \sum_{ia=1}^{N_\text{exc}}
 \int _0^1 \!\! d\alpha\,
 \left\{  \frac{d_{ia}(\i\omega)}{1-\alpha d_{ia}(\i\omega)} - d_{ia}(\i\omega)
 \right\}
 \nonumber \\ & =
 \half
 \int_{-\infty}^{\infty}\!\! \frac{d\omega}{2\pi}\,
 \sum_{ia=1}^{N_\text{exc}}
 \left\{  \log \bigl(1- d_{ia}(\i\omega)\bigr) + d_{ia}(\i\omega)
 \right\}
\nonumber\\&=
\label{eq:dRPAIlog}
   \half \int_{-\infty}^{\infty}\!\! \frac{d\omega}{2\pi}\,
   \tr
   \left\{
   \log \left(\b{I}-\b{{\Pi}}_0(\i\omega )\,\b{K}\right) +
   \b{{\Pi}}_0(\i\omega )\,\b{K}
   \right\}
.\end{align}
We recognize here the matrix representation of the dielectric function,  $\B{\epsilon}(\i\omega)=\b{I}-\b{\Pi}_0(\i\omega){\b{K}}$,
and this expression  may be called the \textit{dielectric matrix formulation} of the dRPA-I correlation energy.

Although it might not be obvious at first sight, a further analytical integration on the frequency
leads directly to the \textit{plasmon formula}.
Such a relationship has  already been mentioned by McLachlan~\textit{et al.} in the sixties~\cite{McLachlan:63}. Inversely, the dielectric matrix formulation can be derived from the plasmon formula, as it has been shown recently by Eshuis \textit{et al.}~\cite{Eshuis:10}.  In Appendix~\ref{app:plasmon}, we present an alternative derivation of the plasmon formula for the dRPA-I correlation energy from the dielectric formulation of Eq.~(\ref{eq:dRPAIsmall}).
This seals the strict equivalence of all formalisms in the case of dRPA-I.
Notice that the plasmon formula applies only
to the dRPA-I (and to the RPAx-II, which is not discussed here) energy expressions.

Moreover, doing a series expansion of the matrix logarithm in Eq.~(\ref{eq:dRPAIlog}) and retaining only the first non-vanishing term, we obtain:

\begin{align}
E_{\text{c}}^{\text{dRPA-I}~(2)}&=
   {-}\frac{1}{4} \int_{-\infty}^{\infty}\!\! \frac{d\omega}{2\pi}\,
   \tr
   \left\{
   \b{{\Pi}}_0(\i\omega )\,\b{K}
   \b{{\Pi}}_0(\i\omega )\,\b{K}
   \right\}
\nonumber\\&=
   {-}\frac{1}{4} \int_{-\infty}^{\infty}\!\! \frac{d\omega}{2\pi}\,
   \sum_{ia,jb}\left\{
    \frac{-2{\epsilon}_{ia}}{{\epsilon}_{ia}^2 + \omega^2}\,{K}_{ia,jb}\,
    \frac{-2{\epsilon}_{jb}}{{\epsilon}_{jb}^2 + \omega^2}\,{K}_{jb,ia}
   \right\}
\nonumber\\&=
   \mhalf
   \sum_{ia,jb}
    \frac{{K}_{ia,jb}\,{K}_{jb,ia}}{{\epsilon}_{ia} +{\epsilon}_{jb}}
=E_{\text{c}}^\text{dMP2}
,\end{align}
which corresponds to a direct-MP2 method, that is to say a version of the MP2 energy without exchange terms~\cite{Furche:01b} (in its range-separated variant, JMP2~\cite{Janesko:09b}).

\subsection{RPA+SOSEX}\label{RPASOSEX}

Considering the fact that, in the  ACFDT expression, the polarization propagator carries information about the screening properties in the system, a SOSEX-like \textit{Ansatz} consists
in replacing the direct interaction matrix  ${\bb{V}}$ by the
antisymmetrized expression ${\bb{W}}$,

\begin{align}\label{eq:Winteraction}
\bb{G}={\bb{W}}=
\begin{pmatrix}
 \b{A}^\prime&   \b{B}  \\
\b{B}& \b{A}^\prime
\end{pmatrix}
,\end{align}
where the antisymmetrized two-electron integral matrices  have the elements $A^\prime_{ia,jb} =\bra{ib}aj\rangle-\bra{ib}ja\rangle$ and $B_{ia,jb} =\bra{ij}ab\rangle-\bra{ij}ba\rangle$. This particular form of the interaction matrix  can be derived from the time dependent Hartree-Fock equations using the fact that the independent particle wave function, providing the bare, unscreened response, satisfies the Brillouin theorem. We note that one can argue for other choices of the ${\bb{W}}$ matrix (\textit{vide infra}).

The resulting correlation energy expression corresponds to the dRPA-II variant~\cite{Angyan:11},

\begin{align}\label{eq:dRPAIIfull}
 E^{\text{dRPA-II}}_{\text{c}}\! =
 \mhalf
 \int _0^1 \!\! d\alpha
 \int_{-\infty}^{\infty}\!\! \frac{d\omega}{2\pi}\,\Tr
 \left\{ \bb{\Pi}_{\alpha } (\i\omega)\,{\bb{W}} -
         \bb{\Pi}_0 (\i\omega)\,{\bb{W}}
 \right\}
,\end{align}
which can be 
designated also as screened exchange dRPA (dRPA-SX), since the dRPA response function screens  the full, Coulomb plus exchange interaction represented by the matrix ${\bb{W}}$.

As a consequence of the more complicated block structure of the $\bb{W}$ interaction matrix, it is less
obvious to reduce the dimensions of the problem in the dRPA-II correlation energy as compared to
the dRPA-I case.
It is reasonable to decompose $\bb{W}$
into a "major" contribution, involving four identical blocks $\b{B}$
and a "minor" correction, which consists in
the $(\b{A}^\prime -\b{B})$ difference of the diagonal blocks:

\begin{align}\label{eq:approxkernel}
\bb{W}=
\begin{pmatrix}
\b{B} & \b{B}\\
\b{B} & \b{B}
\end{pmatrix} +
\begin{pmatrix}
(\b{A}^\prime-\b{B}) & 0\\
0 & (\b{A}^\prime-\b{B})
\end{pmatrix}
.\end{align}

The major contribution to the integrand
can be brought to a convenient form by applying the unitary transformation $\bb{U}$ (Eq.~\ref{eq:unitary})
to the matrix products under the trace in Eq.~(\ref{eq:dRPAIIfull}).
After evaluating the inverse of the
blocked matrix and subsequent matrix multiplications,
one gets the trace of the major contribution in dimension-reduced
matrices.
The minor contribution to the energy expression of Eq.~(\ref{eq:dRPAIIfull}) can be neglected, as is shown in Sec.~2 of the Supporting Information
, and
this leads to
the following approximation of the screened exchange dRPA, denoted in
our previous publication~\cite{Angyan:11} as dRPA-IIa:

\begin{align}\label{eq:dRPAIIafullsmall}
 E^{\text{dRPA-IIa}}_{\text{c}}&=
 \mhalf
 \int _0^1 \!\! d\alpha
 \int_{-\infty}^{\infty}\!\! \frac{d\omega}{2\pi}\,\tr
 \left\{ \b{\Pi}_{\alpha } (\i\omega)\,{\b{B}} -
         \b{\Pi}_0 (\i\omega)\,{\b{B}}
 \right\}
.\end{align}

As in the dRPA-I case, after analytical frequency integration we get the same density matrix formulation expression as the one we have obtained in a quite different manner in Ref.~\onlinecite{Angyan:11}.

Before exploring another alternative, which consists in an analytical integration according to the interaction strength parameter,  the reader must bear in mind the close analogy of dRPA-IIa and the so-called SOSEX (second order screened exchange) approximation~\cite{Gruneis:09,Freeman:77}, usually defined in the framework of a  drCCD (direct ring coupled cluster doubles) theory. Although the above ACFDT-based expression is not strictly equal to the drCCD-based SOSEX, their difference appears only at the third order of perturbation, as was demonstrated in Ref.~\onlinecite{Jansen:10}. The difference of those two variants has been found numerically small for all of the systems studied up to now~\cite{Angyan:11}.
The approximation, which leads from dRPA-II to dRPA-IIa shows in a self-explanatory manner the perturbation character of the SOSEX (dRPA-IIa), \textit{i.e.}\ "second-order screened exchange" compared to  the fully screened exchange (SX, \textit{i.e.}\ dRPA-II) version.

In order to perform the analytical integration along the adiabatic connection path, we 
take advantage of the matrix function formalism (with now obvious notations for $\b{Q}$ and $d_{ia}$):
\begin{widetext}
\begin{align}
\label{eq:dRPAIIeigenv}
E^{\text{dRPA-IIa}}_{\text{c} } & =  
\mhalf
 \int_{-\infty}^{\infty}\!\! \frac{d\omega}{2\pi}\,
 \sum_{ia=1}^{N_\text{exc}}
 \int _0^1 \!\! d\alpha\,
\left\{
 \biggl(
 \bigl(1-\alpha d_{ia}(\i\omega)\bigr)^{-1}-1\biggr)
 \bigl(\b{Q}^{-1}\b{\Pi}_0(\i\omega)\b{B}\b{Q}\bigr)_{ia,ia}
\right\} 
\nonumber\\&=
\half \int_{-\infty}^{\infty}\!\! \frac{d\omega}{2\pi}\,
\sum_{ia=1}^{N_\text{exc}}
\left\{
\biggl(\log\bigl(1-d_{ia}(\i\omega)\bigr)\,d_{ia}^{-1}(\i\omega)+1
\biggr)\bigl(\b{Q}^{-1}\b{\Pi}_0(\i\omega)\b{B}\b{Q}\bigr)_{ia,ia}
\right\} 
\nonumber\\&=
\half \int_{-\infty}^{\infty}\!\! \frac{d\omega}{2\pi}\,
   \tr\left\{
   \log
   \left(\b{I}-\b{\Pi}_0(\i\omega)\b{K}\right)
   \b{K}^{-1}\b{B}  +
   \b{\Pi}_0(\i\omega)\,\b{B}\right\}
.\end{align}
\end{widetext}
Particular care has to be taken to define the inverse of $\b{K}$ in Eq.~(\ref{eq:dRPAIIeigenv}), since this matrix might be singular or close to singular in certain
representations. Since the (close to) zero eigenvalues $d_{ia}(\i\omega )$ do not contribute to the sum over $ia$ in Eq.~(\ref{eq:dRPAIIeigenv}), as can be seen from an expansion of the logarithm, it is implicit
that $\b{K}^{-1}=\b{K}^{-1}\b{\Pi}_0^{-1}\b{\Pi}_0=\b{Q}\b{D}^{-1}\b{Q}^{-1}\b{\Pi}_0$ is defined in terms of a pseudoinverse 
where $[\b{D}^{-1}]_{ia,ia} =d_{ia}^{-1}$ for finite $d_{ia}$ and $[\b{D}^{-1}]_{ia,ia}=0$ for $d_{ia}=0$ (or below a certain small threshold).

By decomposing the antisymmetrized two-electron integrals as $\b{B}=\b{K} - \b{\tilde{K}}$, where  the $\b{\tilde{K}}$ matrix has the elements $ \tilde{K}_{ia,jb} = \bra{ab}ji\rangle$, the
correlation energy can be  separated to a dRPA-I contribution and a SOSEX correction:

\begin{align}\label{eq:RPAIIalarge}
&E_{\text{c}}^{\text{dRPA-IIa}} = E_{\text{c}}^{\text{dRPA-I}} + E_{\text{c}}^\text{SOSEX}
\\
&E_{\text{c}}^\text{SOSEX} =\mhalf
 \int_{-\infty}^{\infty}\!\! \frac{d\omega}{2\pi}\,
 \times
\nonumber\\&\qquad\qquad\;
   \tr
   \left\{\biggr(
   \log
   \left(
   \b{I}-\b{{\Pi}}_0(\i\omega)\b{K}
   \right)  +
   \b{{\Pi}}_0(\i\omega)\,\b{K}\biggr)
   \b{K}^{-1}\b{\tilde{K}}
   \right\}
.\end{align}

Furthermore, it is easy to verify that the second order approximation to the dRPA-IIa correlation energy is exactly the usual MP2 correlation energy
(again by doing a series expansion of the matrix logarithm):

\begin{align}\label{eq:MP2log}
   E_{\text{c}}^{\text{dRPA-IIa}~(2)}  &=
   {-}\frac{1}{4}
   \int_{-\infty}^{\infty}\!\! \frac{d\omega}{2\pi}\,
 \tr\left\{
  \b{{\Pi}}_0(\i\omega) \b{K}\b{{\Pi}}_0(\i\omega)\b{B}
   \right\}
\nonumber\\&=
   \mhalf
   \sum_{ia,jb}
    \frac{{K}_{ia,jb}\,{B}_{jb,ia}}{{\epsilon}_{ia} +{\epsilon}_{jb}}
 = E_{\text{c}}^{\text{MP2}}
.\end{align}

An identical result has been obtained in the Appendix of Ref.~\onlinecite{Angyan:11}.  It is important to mention that in this context the MP2  correlation energy represents the first non-vanishing contribution to a converging series only if all the eigenvalues of $\b{{\Pi}}_0(\i\omega) \b{K}$ are smaller than 1.

\subsection{Approximate RPA with exchange}\label{sec:RPAx}

The RPAx variants are based on a response function
$\bb{\Pi}_\alpha^\text{RPAx}$ which fully takes into account
the nonlocal exchange effects by a Hartree-Fock type kernel. In other words, the
response function $\bb{\Pi}_\alpha^\text{RPAx}$ satisfies the following
Dyson-like equation, where $\bb{F}=\bb{W}$ in Eq.~(\ref{eq:Dysoninv}):

\begin{align}\label{eq:DysoninvRPAx}
   \bb{\Pi}_\alpha^\text{RPAx}(\i\omega)=
   (\bb{I}-\alpha\,\bb{\Pi}_0(\i\omega)\,
    \bb{W})^{-1}\bb{\Pi}_0(\i\omega)
.\end{align}

In the RPAx-I case, that is to say with $\bb{G}=\bb{V}$ in Eq.~(\ref{eq:ACFDTformula}), the ACFDT expression becomes:

\begin{align}\label{eq:RPAx-I}
E^{\text{RPAx-I}}_{\text{c}}&=
 \mhalf
  \int_0^1 \!\! d\alpha
 \int_{-\infty}^{\infty}\!\! \frac{d\omega}{2\pi}\,
 \times 
\nonumber\\&\quad 
   \tr\left\{
   \bigl(
   \mathbb{I}-\alpha\,\mathbb{\Pi}_0(\i\omega)\,\mathbb{W}
   \bigr)^{-1}\,
   \mathbb{\Pi}_0(\i\omega)\mathbb{V}-
   \mathbb{\Pi}_0(\i\omega)\mathbb{V}
   \right\}
.\end{align} 
Let us write again $\mathbb{W}$ as a sum of two terms,

\begin{align}\label{eq:Wintersplit}
{\mathbb{W}}=
\begin{pmatrix}
 \b{B}&   \b{B}  \\
\b{B}& \b{B}     
\end{pmatrix} + 
\begin{pmatrix}
 \b{A}^\prime-\b{B}&   \b{0}  \\
\b{0}& \b{A}^\prime -\b{B} 
\end{pmatrix}
,\end{align}
and use the transformation $\bb{U}$ on the matrix product under the trace.
Neglecting the minor correction contribution leads to the RPAx-Ia energy:

\begin{align}\label{eq:RPAxIfullsmall}
E^{\text{RPAx-Ia}}_{\text{c}} &=
 \mhalf
 \int_0^1 \!\! d\alpha\,
 \int_{-\infty}^{\infty}\!\! \frac{d\omega}{2\pi}\,
   \tr\left\{
\b{\Pi}_\alpha^\text{RPAX}(\i\omega)
\b{K}- 
   \b{\Pi}_0(\i\omega)\b{K}
   \right\}
,\end{align}
where we define $\b{\Pi}_\alpha^\text{RPAX}(\i\omega)$ which satisfies the dimension-reduced Dyson-like equation, obtained  from Eq.~(\ref{eq:DysoninvRPAx}) after taking an approximate kernel  $\mathbb{F}^\text{RPAX}=\mathbb{B}$ including only the major contribution from Eq.~(\ref{eq:Wintersplit}):

\begin{align}
\b{\Pi}_\alpha^\text{RPAX}(\i\omega)=\bigl(
   \b{I}-\alpha \b{\Pi}_0(\i\omega)\b{B}
   \bigr)^{-1}\,
   \b{\Pi}_0(\i\omega)
.\end{align}
The notation RPAX (with capital X) refers to an analogy to an approximate RPA variant with exchange, proposed by Hesselmann~\cite{Hesselmann:12}. The relationships to Hesselmann's approach will be discussed elsewhere.
Comparison to Eqs.~(\ref{eq:dRPAIIafullsmall}) and~(\ref{eq:Dysoninvsmall}) shows that, formally, the roles of the matrices $\b{K}$ and $\b{B}$ is merely exchanged with respect to the dRPA-IIa energy expression.
Hence, the corresponding logarithmic formula obtained after $\alpha$-integration is:

\begin{align}\label{eq:RPAxIlog}
E^{\text{RPAx-Ia}}_{\text{c}}& =
 \half
 \int_{-\infty}^{\infty}\!\! \frac{d\omega}{2\pi}\,
   \tr\left\{
   \log\bigl(
   \b{I}-
   \b{\Pi}_0(\i\omega)\b{B}
   \bigr)\,\b{B}^{-1}\b{K}+
   \b{\Pi}_0(\i\omega)\b{K}
   \right\}
,\end{align}
again defining $\b{B}^{-1}$ in terms of a pseudoinverse involving the eigenvalues of $\b{\Pi}_0(\i\omega)\b{B}$.

The second order approximation yields the MP2 energy in the exact same manner as the dRPA-IIa case (consider the exchange of the matrices $\b{K}$ and $\b{B}$ in Eq.~(\ref{eq:MP2log})). 

As shown in Sec.~\ref{numerical}, the approximate RPAx-Ia correlation energy expression leads to the most accurate results for the systems
considered in this work. Additionally, while the RPAx-I approximation (Eq.~\ref{eq:RPAx-I}) is known to suffer from instabilities when
the initial states are approximated from semi-local functionals~\cite{Angyan:11},  
 the RPAx-Ia is numerically stable.

\section{Computational realization}\label{computational}

For the practical implementation, it is worth noting that spin-adaptation of all the equations is trivial: in spacial orbitals, the two-electron integrals read as $K_{ia,jb}=\!\! = \!\!2\bra{ij} ab\rangle$, $A^\prime_{ia,jb}\!\! = \!\!2\bra{ib} aj\rangle-\bra{ib} ja\rangle$, $B_{ia,jb}\!\! = \!\!2\bra{ij} ab\rangle-\bra{ij} ba\rangle$ and $\tilde{K}_{ia,jb}\!\! = \!\!\bra{ij} ba\rangle$.

In the following, we present an orbital-based implementation of the equations derived in Sec.~\ref{sec:RPA}. An implementation using density fitting is shown in the Appendix.~\ref{app:DF}.

\subsection{Orbital-based implementation}

The computational realization of equations~(\ref{eq:dRPAIlog}), (\ref{eq:dRPAIIeigenv}) and (\ref{eq:RPAxIlog}) may proceed  following a common scheme. In fact, Eq.~(\ref{eq:dRPAIlog}) can be considered as a special case of Eq.~(\ref{eq:dRPAIIeigenv}), where  $\b{B}=\b{K}$, and   Eq.~(\ref{eq:RPAxIlog}) can be obtained by interchanging the roles of $\b{B}$ and $\b{K}$ in Eq.~(\ref{eq:dRPAIIeigenv}). Therefore we focus our attention to the case of Eq.~(\ref{eq:dRPAIIeigenv}), \textit{i.e.}\ the dRPA-IIa energy expression.

We can rewrite Eq.~(\ref{eq:dRPAIIeigenv}) in terms of the symmetric matrices $\b{P}(\i\omega)=\b{{\Pi}}_0^{1/2}(\i\omega)\b{K}\b{{\Pi}}_0^{1/2}(\i\omega)$ and  $\b{\tilde{P}}(\i\omega)=\b{{\Pi}}_0^{1/2}(\i\omega)\b{\tilde{K}}\b{{\Pi}}_0^{1/2}(\i\omega)$ as

\begin{align}
   E_{\text{c}}^{\text{dRPA-IIa}} &=
   \int_0^{\infty}\!\! \frac{d\omega}{2\pi}\,
   \tr\biggl\{\bigl(
   \log\bigl(\b{I}-
   \b{P}(\i\omega)\bigr)+\b{P}(\i\omega)\bigr) \times
   \nonumber\\&\qquad\qquad\qquad\qquad\qquad 
   \bigl(\b{I}-
   \b{P}^{-1}(\i\omega)\,
   \b{\tilde{P}}(\i\omega)\bigr)\biggr\}
\label{eq:intermediate2}
,\end{align}
where we took advantage of the cyclic invariance of the trace and used the symmetry of $\b{\Pi}_0(\i\omega)$ with respect to the replacement of $\i\omega$ by $-\i\omega$ to restrain the integral to the positive imaginary axis.
The derivation of Eq.~(\ref{eq:intermediate2}) also requires
the following identity for a generic function of a matrix $f$:

\begin{align}
f(\b{{\Pi}}_0^{1/2}\b{K}\b{{\Pi}}_0^{1/2}) = 
\b{{\Pi}}_0^{-1/2} f(\b{{\Pi}}_0\b{K}) \b{{\Pi}}_0^{1/2}
,\end{align}
that can be easily derived from the matrix function formalism and from the fact that $\b{Q}^{-1} \b{{\Pi}}_0 \b{K} \b{Q}=\b{D}$
implies $(\b{Q}^{-1} \b{{\Pi}}_0^{1/2}) \b{{\Pi}}_0^{1/2} \b{K} \b{{\Pi}}_0^{1/2} ( \b{{\Pi}}_0^{-1/2} \b{Q}) = \b{U}^\T \b{{\Pi}}_0^{1/2} \b{K} \b{{\Pi}}_0^{1/2} \b{U}  =\b{D}$
(the matrix $\b{U}$ defines the unitary transformation that diagonalizes the Hermitian matrix $\b{{\Pi}}_0^{1/2} \b{K} \b{{\Pi}}_0^{1/2}$).

By expressing Eq.~(\ref{eq:intermediate2}) in terms of the (diagonal) eigenvalue matrix $\b{D}$ we obtain

\begin{align}
   E_{\text{c}}^{\text{dRPA-IIa}} &=
   \int_0^{\infty}\!\! \frac{d\omega}{2\pi}\,
   \tr\biggl\{\bigl(
   \log\bigl(\b{I}-
   \b{D}(\i\omega)\bigr)+\b{D}(\i\omega)\bigr) \times
   \nonumber\\&\qquad\qquad\qquad\qquad\quad
   \bigl(\b{I}-
   \b{D}^{-1}(\i\omega)\,
   \b{\tilde{D}}(\i\omega)\bigr)\biggr\}
,\end{align}
with $\b{\tilde{D}}(\i\omega) = \b{U}^\T\b{\tilde{P}}(\i\omega)\b{U}= \bigl(\b{U}^\T\b{{\Pi}}_0^{1/2}(\i\omega)\bigr)\,\b{\tilde{K}}\,\bigl(\b{{\Pi}}_0^{1/2}(\i\omega)\b{U}\bigr)=\b{Q}^{-1}\b{\Pi}_0(\i\omega)\b{\tilde{K}}\b{Q}$.
It should be emphasized that the matrix $\b{\tilde{P}}$ has been transformed with the eigenvectors of $\b{P}$ and therefore is not diagonal. Nevertheless in order to calculate the trace of 
its product with the diagonal matrix $\b{D}^{-1}$ we need only its diagonal elements, which will be designated by the shorthand notation $\tilde{d}_{ia}(\i\omega)=[\b{\tilde{D}}(\i\omega)]_{ia,ia}=[\b{U}^\T \b{\tilde{P}}(\i\omega)\b{U}]_{ia,ia}$.

The presence of potentially very
small eigenvalues may lead to numerical instabilities.
This can be avoided by using the following power series expansion of the logarithm under a given threshold of the eigenvalue $d_{ia}(\i\omega)$ (in practice, the summation is carried up to $n=4$ and the threshold is 0.0001):

\begin{align}
   E_{\text{c}}^{\text{dRPA-IIa}}\approx
   -\int_0^\infty\frac{d\omega}{2\pi}\,
   \sum_{ia=1}^{N_\text{exc}}
   \sum_{n=2}
      \frac{1}{n} d^{n-1}_{ia}(\i\omega)\,
   \left(d_{ia}(\i\omega)-\tilde{d}_{ia}(\i\omega)\right)
.\end{align}

At second order, $n=2$, we obtain the MP2 energy, which can be designated as "Casimir-Polder transform MP2"

\begin{align}
   E_{\text{c}}^{\text{MP2}}=
   \mhalf \int_0^\infty\frac{d\omega}{2\pi}\,
   \sum_{ia}^{N_\text{exc}}
      d_{ia}(\i\omega)\,
   \left(d_{ia}(\i\omega)-\tilde{d}_{ia}(\i\omega)\right)
,\end{align}
and which offers
an interesting alternative to calculate conventional MP2 energies, especially in solids~\cite{Marsman:09}.
Higher order terms of the series expansion do not correspond exactly to the MP3 and MP4 energies.
A vague analogy can be noted between this expression and the Laplace-transform method to obtain the MP2 energy~\cite{Eshuis:10}.

\subsection{Numerical frequency integration}\label{sec:numerical}

In principle, the numerical frequency integration is expected to be "fairly unproblematic"~\cite{Harl:10},
since the integrand is expected to have a smoothly decaying behavior.
While it seems to be really the case for solids, where a
mapping of the Gauss-Legendre quadrature to the $[0,\infty]$ interval (truncated at about 30~a.u.)
with an exponentially decaying weighting function ensures a reasonable convergence (0.05~mH) with only 16 quadrature points~\cite{Harl:10},
an accuracy of 0.2~mH (about 0.15~kcal/mol) claimed by the same authors  is clearly insufficient
for atoms and molecules.
The situation for atomic and molecular systems is not better for more sophisticated weighting function models either.

Lu~\textit{et al.} have performed the frequency integration  by a 10-point Gauss-Legendre quadrature in the range of $u\in[0,1]$, where $u=(1+\omega/\omega_0)^{-1}$, with $\omega_0=1$~a.u. This quadrature
was found rather insensitive to the choice of the $\omega_0$ parameter~\cite{Lu:09}.

We have implemented the method proposed by Eshuis, Yarkony and Furche~\cite{Eshuis:10}, based on the Clenshaw-Curtis quadrature~\cite{Boyd:87}, which consists in writing the integral of a function $F(\omega)$ as

\begin{align}
  \int_0^{\infty}\!\! d\omega\, F(\omega) \approx
  \sum_p \frac{w_p}{2\pi}\,F(\omega_p)
,\end{align}
where $\omega_p = a \cot t_p$ with $t_p = p \pi /(2 N_g )$   and $w_p = a \,\pi /((\text{int}(p/N_g)+1) N_g \text{sin}^2 t_p)$.
This quadrature scheme has a single scaling parameter $a$, which
can be adjusted to each individual system
by requiring that the following analytically solvable expression, based on a diagonal model of the dRPA dielectric matrix formulation of Eq.~(\ref{eq:dRPAIlog})
(remember that $\b{\Pi}_0(\i\omega)=-2\B{\epsilon}^{1/2}(\B{\epsilon}^2+\omega^2)^{-1}\B{\epsilon}^{1/2}$)
:

\begin{align}\label{eq:dRPAIdiag}
   E_{\text{c}}^{\text{dRPA-I}} &\approx
   \half \sum_{ia} \int_{-\infty}^{\infty}\!\! \frac{d\omega}{2\pi}\,
   \left\{
   \log \left(1 +
   \frac{2\epsilon_{ia}K_{ia,ia}}
        {\epsilon_{ia}^2 + \omega^2}
   \right)-
   \frac{2\epsilon_{ia}K_{ia,ia}}
        {\epsilon_{ia}^2 + \omega^2}
   \right\}
   \nonumber \\ & = \mhalf\sum_{ia}
   \biggl(
   \epsilon_{ia} + K_{ia,ia} - \sqrt{\epsilon^2_{ia}+2\epsilon_{ia}K_{ia,ia}}
   \biggr)
,\end{align}
be reproduced at best by the  numerical integral,

\begin{align}
E_{\text{c}}^{\text{dRPA-I}}&\approx
   \sum_{ia}
   \sum_p
   \frac{w_p}{2\pi}\,
\biggl\{\log\biggl(1+
   \frac{2\epsilon_{ia}K_{ia,ia}}
        {\epsilon_{ia}^2+\omega_p^2} \biggr)-
   \frac{2\epsilon_{ia}K_{ia,ia}}
        {\epsilon_{ia}^2+\omega_p^2} \biggr\}
.\end{align}

This tuning method of the parameter $a$ can be extended for the dRPA-IIa case,
using the analytically solvable dRPA-IIa diagonal approximation:

\begin{align}\label{eq:dRPAIIadiag}
   E_{\text{c}}^{\text{dRPA-IIa}} \approx &
   \half \sum_{ia} \int_{-\infty}^{\infty}\!\! \frac{d\omega}{2\pi}\,
   \left\{
   \log \left(1 +
   \frac{2\epsilon_{ia}K_{ia,ia}}
        {\epsilon_{ia}^2 + \omega^2}
   \right)-
   \frac{2\epsilon_{ia}K_{ia,ia}}
        {\epsilon_{ia}^2 + \omega^2}
   \right\}
\nonumber\\&\qquad\qquad\qquad\times\left\{
   1-
\frac{{\tilde{K}}_{ia,ia}}
     {{K}_{ia,ia}}
   \right\}
   \nonumber \\  = & {-}\frac{1}{4}\sum_{ia}
   \biggl\{
   \biggl(
   \epsilon_{ia} + K_{ia,ia} - \sqrt{\epsilon_{ia}^2 +2 \epsilon_{ia}K_{ia,ia}}
   \biggr)
   \biggr\}
.\end{align}
Note that the diagonal approximation to the dRPA-IIa correlation energy
is half the diagonal approximation to the dRPA-I correlation energy seen in Eq.~\ref{eq:dRPAIdiag}
(remember that in spatial orbitals we have  $K_{ia,jb}\!\! = \!\!2 \bra{ij} ab\rangle$ and $\tilde{K}_{ia,jb}\!\! = \!\!\bra{ij} ba\rangle$)
and there is no need
to re-optimize the free parameter of the numerical quadrature.

\section{Numerical results}\label{numerical}

We have implemented the dielectric matrix based {dRPA-I} (Eq.~\ref{eq:dRPAIlog}), dRPA-IIa (Eq.~\ref{eq:dRPAIIeigenv}), and RPAx-Ia (Eq.~\ref{eq:RPAxIlog}) 
energy formulae in an occupied-virtual
basis set representation within the development version of the MOLPRO quantum chemistry package~\cite{MOLPRO_brief,MOLPRO-WIREs}.
Although this algorithm is expected to be less efficient
than the density fitting approach (in particular for larger systems), we considered this implementation useful to produce benchmark results 
exempt of density fitting uncertainties.  

Below we present electron correlation energies for some atoms and ions~\cite{davidson1,davidson2} and a test set of reaction energies~\cite{Hesselmann:12}.
The correlation energy from the different RPA approximations $E_{\text{c}}^\text{RPA}$ 
has been calculated starting from a self-consistent DFT calculation based on
the Perdew-Burke-Ernzerhof (PBE) exchange-correlation functional. The total RPA energy is then evaluated 
from the expression
\begin{equation}
E_{\text{tot}}^{\text{RPA}}=E_{\text{EXX}}+E_{\text{c}}^{\text{RPA}},
\end{equation}
where $E_{\text{EXX}}$ is the Hartree-Fock energy computed from PBE orbitals.

For comparison purposes, we also provide results obtained from the rCCD-based
SOSEX approximation~\cite{Gruneis:09,Freeman:77} and the dRPA-II approximation as derived within the
adiabatic connection approach~\cite{Angyan:11}. As already discussed in Sec.~\ref{sec:introduction} and Sec.~\ref{RPASOSEX} the
dRPA-IIa (Eq.~\ref{eq:dRPAIIeigenv}) and the rCCD-based SOSEX are analogous although not strictly equivalent;
the numerical calculations below help to better quantify this statement. The dRPA-II approximation as derived within the
adiabatic connection approach is equivalent to the dynamical polarizability expression in Eq.~\ref{eq:dRPAIIfull},
that, however, cannot be conveniently implemented (both frequency and coupling
constant integration are necessary and the dimensionality of the polarizability cannot be reduced).
In this case the comparison between the dielectric matrix-based dRPA-IIa and the adiabatic connection-based dRPA-II
is useful to understand the effect of the approximation of the kernel introduced by discarding the minor contribution to Eq.~\ref{eq:approxkernel}. 
To avoid confusion the dRPA-I, dRPA-IIa, and RPAx-Ia approximations derived within the dielectric matrix
formulation will be denoted with an additional ``DIEL'', the SOSEX approximation with an additional ``rCCD'', and
the dRPA-II approximation within the adiabatic connection approach with an additional ``AC''.

\subsection{Frequency quadrature}

\begin{figure}
  \includegraphics[trim=0mm 0 0mm 0,clip=true,height=60mm,keepaspectratio]{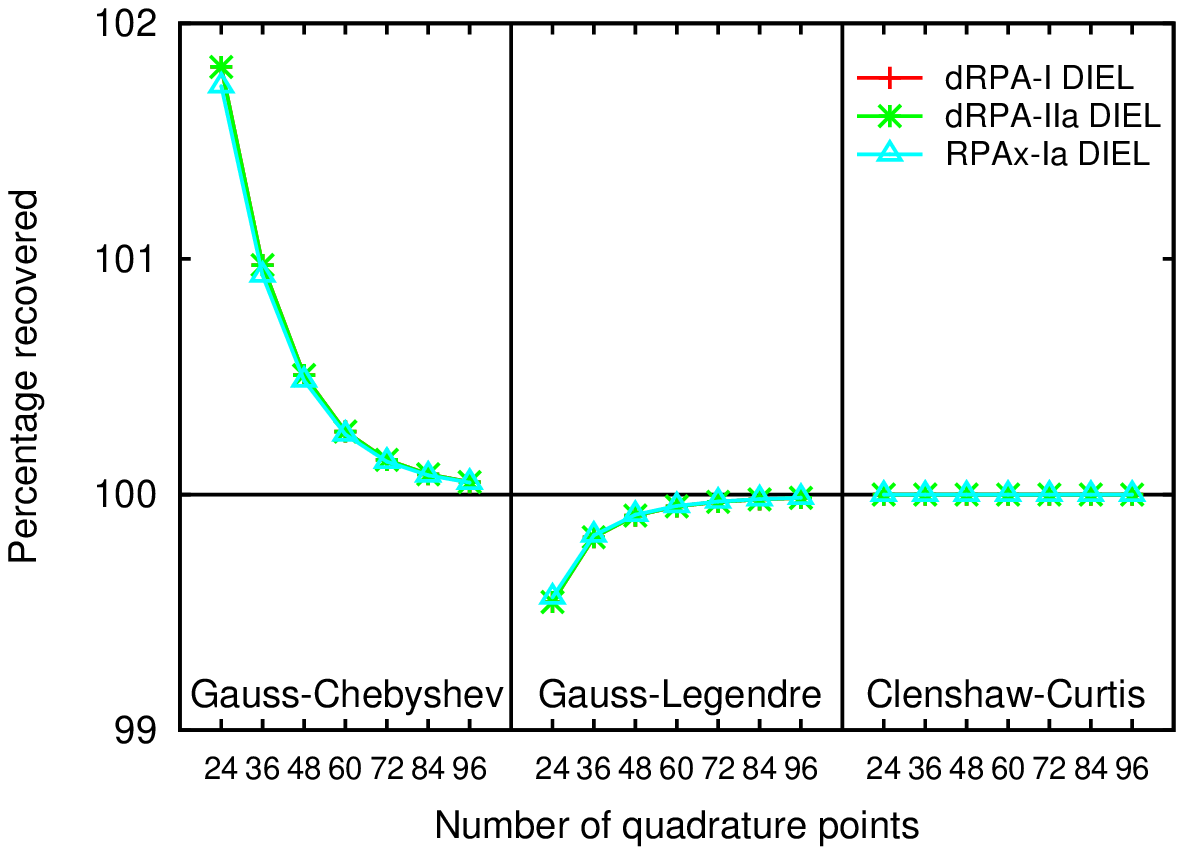}

  \includegraphics[trim=0mm 0 0mm 0,clip=true,height=60mm,keepaspectratio]{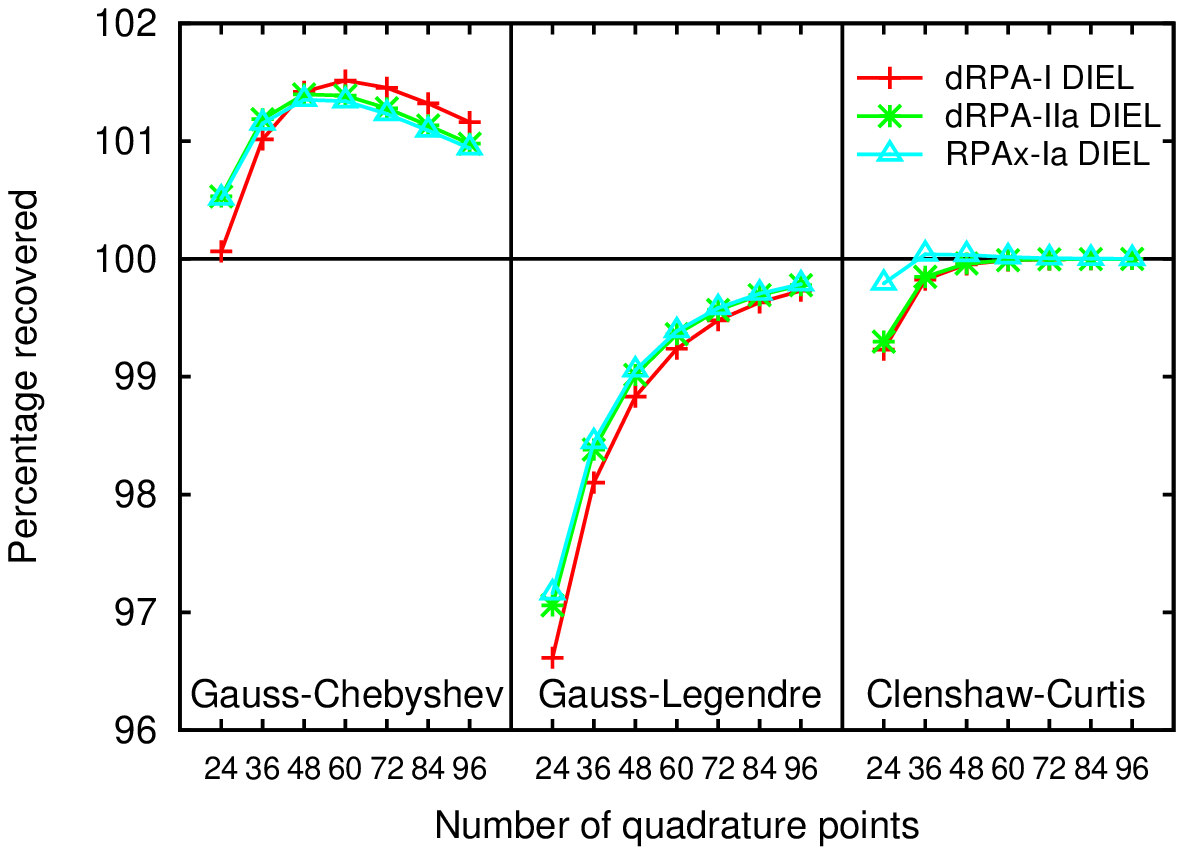}
  \caption{Convergence with respect to the number of quadrature points of the Gauss-Chebyshev, Gauss-Legendre and Clenshaw-Curtis schemes for Li$^+$ (up) and Mg (down).
The dRPA-I DIEL energy is compared to the dRPA-I PLASMON energy, and the dRPA-IIa DIEL and RPAx-Ia DIEL are compared against their AC analog.} \label{convgraph}
\end{figure}

The frequency integrations in Eqs.~\ref{eq:dRPAIlog},~\ref{eq:dRPAIIeigenv} and ~\ref{eq:RPAxIlog} are carried out by a quadrature.
We tested the convergence of the Gauss-Chebyshev\cite{Rijks:88}, Gauss-Legendre\cite{Furche:01b} and Clenshaw-Curtis
(described in Sec.~\ref{sec:numerical})
schemes
with respect to the number of quadrature points,
for calculations on several atoms and ions.
The RPA correlation energies were computed using the aug-cc-pCVXZ basis sets (X=6 for He, X=5 for Ne, Ar, B$^+$, and Al$^+$, and X=Q for Li$^+$, Na$^+$, Be, and Mg). 
The results of the convergence study can be found in Table~\ref{tab:conv} and Figure~\ref{convgraph}, where we use as reference
the dRPA-I PLASMON (for which both the frequency and coupling constant integrations are analytical, see Appendix~\ref{app:plasmon}), and the dRPA-IIa AC and RPAx-Ia AC energies (for which no PLASMON analog exists).
We see that the Gauss-Chebyshev and Gauss-Legendre schemes yield, in this case, unsatisfying results:
the Gauss-Chebyshev scheme has a slow convergence with respect to the number of quadrature points in all the cases studied,
and while the Gauss-Legendre quadrature performs well for Be, He and Li$^+$, it exhibits a slow convergence in the cases of B$^+$, Mg and Na$^+$.
On the other hand, the Clenshaw-Curtis quadrature was found to converge rapidly for all studied atoms and ions.
As a result, we choose in the following to perform the frequency integration with a 48-points
Clenshaw-Curtis quadrature.

\begin{table}
\caption{\label{tab:conv} Errors made in the calculation of the dRPA-I DIEL energy using 24-, 48-, 72- and 96-points Gauss-Chebyshev, Gauss-Legendre and Clenshaw-Curtis quadrature schemes (in percentage with respect to the dRPA-I PLASMON result).}
\begin{tabularx}{\linewidth}{lrrrrrr}
\hline\hline
        &       B+      &       Mg      &       Na+     &       Be      &       He      &       Li+     \\      
\hline
&\multicolumn{6}{c}{Gauss-Chebyshev}\\
24      &        1.51   &        0.07   &        0.90   &        1.63   &        0.51   &        1.81   \\
48      &        1.09   &        1.42   &        1.05   &        0.84   &        0.16   &        0.51   \\
72      &        0.78   &        1.45   &        0.81   &        0.44   &        0.04   &        0.15   \\
96      &        0.58   &        1.16   &        0.66   &        0.22   &        0.01   &        0.05   \\[.5em]
&\multicolumn{6}{c}{Gauss-Legendre}\\
24      &       -2.22   &       -3.38   &       -2.23   &       -1.06   &       -0.13   &       -0.46   \\
48      &       -0.78   &       -1.17   &       -0.78   &       -0.23   &       -0.02   &       -0.09   \\
72      &       -0.37   &       -0.52   &       -0.35   &       -0.09   &       -0.01   &       -0.03   \\
96      &       -0.21   &       -0.27   &       -0.19   &       -0.04   &        0.00   &       -0.01   \\[.5em]
&\multicolumn{6}{c}{Clenshaw-Curtis}\\
24      &       -0.35   &       -0.77   &        0.02   &       -0.23   &        0.00   &        0.00   \\
48      &       -0.02   &       -0.04   &        0.00   &       -0.01   &        0.00   &        0.00   \\
72      &        0.00   &        0.00   &        0.00   &        0.00   &        0.00   &        0.00   \\
96      &        0.00   &        0.00   &        0.00   &        0.00   &        0.00   &        0.00   \\
\hline\hline
\end{tabularx}
\end{table}

\subsection{Atomic correlation energies}\label{atomic}

\begin{figure}
\vspace{0.3cm}
  \includegraphics[width=\columnwidth]{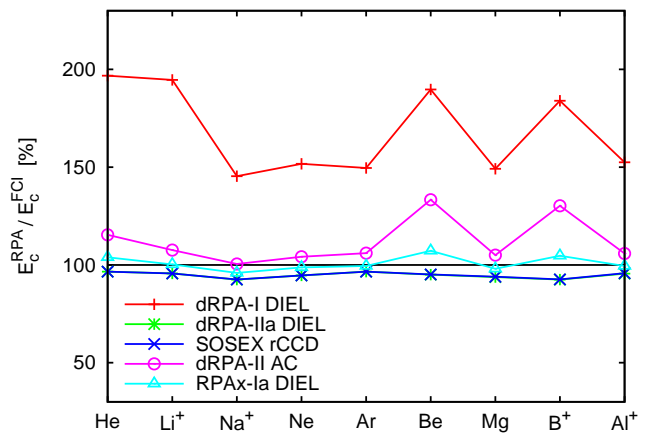}
  \caption{Percentage error of total correlation energies compared to the
FCI-quality estimates by Davidson and his coworkers~\cite{davidson1,davidson2} for simple closed shell atoms and ions calculated with different RPA variants. Correlation energies have been extrapolated to the CBS limit. 
} \label{ecatoms}
\end{figure}

As a first test, we applied the dielectric matrix formalism to compute
correlation energies for several atoms and ions. To verify the
accuracy of the RPA-based approximations we compared the obtained results
with the full configuration interaction (FCI) quality correlation energy estimates by Davidson and
collaborators~\cite{davidson1,davidson2}. For a meaningful comparison it is necessary to keep 
into account that the FCI-quality correlation energies have been obtained with respect to
a Hartree-Fock reference point. For this reason we redefined the RPA correlation energies 
$\tilde{E}_{\text{c}}^{\text{RPA}}$ as the difference of RPA total energies and regular Hartree-Fock energies.
This procedure was already used in Ref.~\onlinecite{Angyan:11}.
Core excitations have been included in the calculations. 
The RPA-based correlation energies were computed in the complete basis
set (CBS) limit by the usual 1/X$^3$ formula~\cite{kutzelnigg1992rates} considering aug-cc-pCVXZ basis sets up to X=6 for He, up
to X=5 for Ne, Ar, B$^+$, and Al$^+$, and up to X=Q for Li$^+$, Na$^+$,
Be, and Mg. 
The ratios between the RPA-based correlation energies $\tilde{E}_{\text{c}}^{\text{RPA}}$ and reference
FCI correlation energies $E_{\text{c}}^{\text{FCI}}$ are shown in Fig.~\ref{ecatoms}. As expected the
dRPA-I DIEL approximation strongly overestimates the absolute value
of the correlation energy (by 50 to 100$~\%$). On the other hand, the dRPA-IIa DIEL and RPAx-Ia DIEL approximations, 
which include exchange contributions, significantly improve over the direct RPA results and lead
to a percentage of the correlation energy close to $100~\%$. On the scale of Fig.~\ref{ecatoms}
the dRPA-IIa DIEL and SOSEX rCCD results are basically indistinguishable (the largest difference is found for Al$^+$,
where SOSEX rCCD gives 95.79$\%$ of the FCI correlation energy and dRPA-IIa 95.57$\%$). 
The dRPA-II AC, which does not rely on the neglect of the minor contribution to the kernel discussed in Eq.~\ref{eq:approxkernel},
provides the worst results among the approximations including exchange effects. 
Additionally, in order to quantify the one-electron self-interaction error, we computed 
correlation energies for the hydrogen atom. Also in this case the exchange contribution significantly improves
the results: The dRPAI-I correlation energy of -0.08 Ha decreases to about -0.04 Ha 
for the dRPA-II, dRPA-IIa, SOSEX, and RPAx-Ia approximations. As expected,                                 
the dRPA-IIa correlation energy was found to be exactly half of the dRPA-I value. 

\subsection{Application to reaction energies}

In this section we discuss a series of results for the reaction energy
test set proposed by Hesselmann~\cite{Hesselmann:12}. In this case 
the accuracy is evaluated with respect to CCSD(T) benchmark results; CCSD values 
are also provided for comparison purposes. 
Similarly to Ref.~\onlinecite{Hesselmann:12}, the correlation energy
in the complete basis set limit is obtained by extrapolating
the values obtained with the aug-cc-pVTZ and aug-cc-pVQZ basis sets.
Reaction energy results are detailed in Table~\ref{tablereact} for the different approximations considered in this work.
The simple dRPA-I DIEL approximation gives substantial mean error (ME) and mean absolute error (MAE): 2.18 kcal/mol and 2.32 kcal/mol, respectively.
The dRPA-IIa DIEL approach, with a MAE of 1.99 kcal/mol and a ME of -1.43 kcal/mol, improves 
to a certain extent the results of dRPA-I DIEL. Table~\ref{tablereact} also shows that dRPA-IIa and SOSEX produce similar results, with
a difference of at most 0.2 kcal/mol. The use 
of the full exchange kernel (Eq.~\ref{eq:approxkernel}), as done in dRPA-II AC, deteriorates the accuracy and leads to a MAE that is even larger than 
in the case of the simple dRPA-I approach.
We finally consider the RPAx-Ia DIEL approximation. The corresponding MAE and ME are significantly decreased with respect to the other methods
 and close to CCSD values.
Considering these results for reaction energies and the previous results for atoms and ions in Sec.~\ref{atomic}, we can 
notice that the RPAx-Ia approach is the most promising within the dielectric
matrix approximations studied in this work. Additionally, 
RPAx-Ia is more stable than RPAx-I (Eq.~\ref{eq:RPAx-I}) when using PBE as starting point~\cite{Angyan:11}.
It is important to notice that the PBE and the non-self-consistent PBE0 (post-PBE) approximations lead to the significant MAEs of 4.68 and 4.57 kcal/mol, respectively.

\begin{turnpage}
\begin{table*}
\caption{\label{tablereact} Reaction energies for 16 chemical reactions (in kcal/mol). The mean error (ME) and mean absolute error (MAE)
are computed with respect to CCSD(T)/CBS. The results with the smallest deviation from the benchmark value are marked in bold face.}
\begin{footnotesize}
\begin{tabularx}{\linewidth}{lrrrrrrrrr}
\hline
\hline
\makebox[0.07\textwidth][l]{Reaction}             & \makebox[0.07\textwidth][c]{dRPA-I}
                                                                    & \makebox[0.07\textwidth][c]{dRPA-IIa}
                                                                              & \makebox[0.07\textwidth][c]{SOSEX}
                                                                                                & \makebox[0.07\textwidth][c]{dRPA-II}
                                                                                                                  & \makebox[0.07\textwidth][c]{RPAx-Ia}
                                                                                                                                     & \makebox[0.07\textwidth][c]{PBE}
                                                                                                                                               & \makebox[0.07\textwidth][c]{PBE0}
                                                                                                                                                        & \makebox[0.07\textwidth][c]{CCSD}
                                                                                                                                                                          & \makebox[0.07\textwidth][c]{CCSD(T)} \\
                                                 & \makebox[0.07\textwidth][c]{DIEL}
                                                                    & \makebox[0.07\textwidth][c]{DIEL}
                                                                              & \makebox[0.07\textwidth][c]{rCCD}
                                                                                                & \makebox[0.07\textwidth][c]{AC}
                                                                                                                  & \makebox[0.07\textwidth][c]{DIEL}
                                                                                                                                     &
                                                                                                                                               & \makebox[0.07\textwidth][c]{post-PBE}
                                                                                                                                                        &                 &        \\
\hline
C$_2$H$_2$+H$_2$  $\rightarrow$  C$_2$H$_4$      &  -48.03          & -51.97  & -51.99          & -48.29          & -51.57           & -52.15           & -55.32          & \textbf{-50.46} & -49.44 \\
C$_2$H$_4$+H$_2$  $\rightarrow$ C$_2$H$_6$       &  -37.54          & -42.41  & -42.23          & -38.31          & \textbf{-40.26}  & -40.66           & -43.96          & -40.49          & -39.47 \\
C$_2$H$_6$+H$_2$  $\rightarrow$  2CH$_4$         &  -17.50          & -19.00  & -18.98          & \textbf{-18.15} & -18.45           & -18.61           & -19.01          & -18.76          & -18.18 \\
CO+H$_2$          $\rightarrow$ HCHO             &   -3.07          &  -4.94  &  -4.98          &  -1.39          &  -4.84           & -12.45           & -12.62          &  \textbf{-5.61} &  -5.47 \\
HCHO+H$_2$        $\rightarrow$  CH$_3$OH        &  -26.88          & -32.75  & -32.55          & -27.77          & -30.96           & \textbf{-29.66}  & -33.44          & -30.78          & -29.70 \\
H$_2$O$_2$+H$_2$  $\rightarrow$  2H$_2$O         &  -82.53          & -92.66  & -92.48          & -86.07          & -89.62           & -81.81           & \textbf{-86.78} & -89.55          & -87.63 \\
C$_2$H$_2$+H$_2$O $\rightarrow$  CH$_3$CHO       &  -38.01          & -39.68  & -39.73          & -37.33          & -39.28           & -44.25           & -45.49          & \textbf{-38.53} & -38.28 \\
C$_2$H$_4$+H$_2$O $\rightarrow$  C$_2$H$_5$OH    &  -13.15          & -15.73  & -15.58          & -12.77          & \textbf{-14.35}  & -15.67           & -17.85          & -14.43          & -14.12 \\
CH$_3$CHO+H$_2$   $\rightarrow$ C$_2$H$_5$OH     &  -23.17          & -28.02  & -27.85          & -23.73          & -26.64           & -23.57           & -27.67          & \textbf{-26.36} & -25.28 \\
CO+NH$_3$         $\rightarrow$  HCONH$_2$       &   -6.61          &  -9.77  &  -9.83          &  -5.06          & \textbf{-9.35}   & -21.04           & -19.80          &  -9.35          & -10.26 \\
CO+H$_2$O         $\rightarrow$ CO$_2$+H$_2$     &   \textbf{-5.04} &  -3.81  &  -3.85          &  -1.93          & -3.80            & -17.62           & -12.86          &  -3.76          &  -6.18 \\
HNCO+NH$_3$       $\rightarrow$ NH$_2$CONH$_2$   &  -17.34          & -22.89  & -22.81          & -18.71          & \textbf{-22.02}  & -18.45           & -23.03          & -22.04          & -20.70 \\
CO+CH$_3$OH       $\rightarrow$  HCOOCH$_3$      &  -10.07          & -12.49  & \textbf{-12.56} &  -8.54          & -12.14           & -22.44           & -20.85          & -12.12          & -13.59 \\
HCOOH+NH$_3$      $\rightarrow$ HCONH$_2$+H$_2$O &   -0.79          &  -1.76  &  -1.76          &  \textbf{-1.17} &  -1.60           &  -1.76           &  -2.03          &  -1.24          &  -1.16 \\
CO+H$_2$O         $\rightarrow$ CO$_2$+H$_2$O    &  -87.57          & -96.47  & -96.33          & -87.99          & \textbf{-93.42}  & -99.43           & -99.64          & -93.31          & -93.81 \\
H$_2$CCO+HCHO     $\rightarrow$ C$_2$H$_4$O+CO   &   -4.99          &  -5.67  &  -5.47          &  -5.44          &  -5.26           &   5.06           &   0.52          &  \textbf{-4.93} &  -3.83 \\
\hline
                                              ME &  2.18            &   -1.43 &  -1.37          &   2.15          &  -0.40           &  -2.34           &  -3.92          &  -0.29          &        \\
                                             MAE &  2.32            &    1.99 &   1.90          &   2.36          &   1.12           &   4.68           &   4.57          &   0.95          &        \\
\hline
\hline
\end{tabularx}
\end{footnotesize}
\end{table*}
\end{turnpage}

The reaction test set in Table~\ref{tablereact} involves only energy differences between molecular
energies. To better understand the performance of each method it is interesting
to also discuss the total energy of each molecule. Detailed results are summarized in Table~\ref{tabletot}, where
the deviation of the total energy of each molecule from CCSD(T) reference values is presented.
Since an error cancellation is expected when energy differences are computed, it is not surprising that the approximations
in Table~\ref{tabletot} lead to MEs and MAEs with respect to CCSD(T) that are substantially larger than what previously seen in Table~\ref{tablereact}.
For total energies the dRPA-I approximation leads to the largest deviation with a MAE of about 168 kcal/mol.
The dRPA-IIa DIEL, the SOSEX rCCD, and the dRPA-II AC approximations all have a considerably lower MAE (about 30 kcal/mol).
However, it is important to notice that dRPA-II AC tends to overestimate the total energy while dRPA-IIa DIEL and SOSEX rCCD underestimate it.
Considering Table~\ref{tabletot}, the method that gives the most accurate results is RPAx-Ia DIEL, which provides a MAE of less than
9 kcal/mol and outperforms CCSD.

\begin{figure}
\vspace{0.3cm}
  \includegraphics[width=0.9\columnwidth]{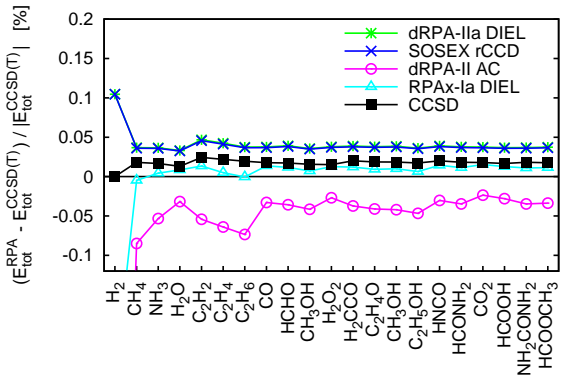}
  \caption{Relative error of the total energy computed with different RPA methods with respect to CCSD(T)/CBS results.} \label{totalenergymol}
\end{figure}

\begin{figure}
  \includegraphics[width=0.9\columnwidth]{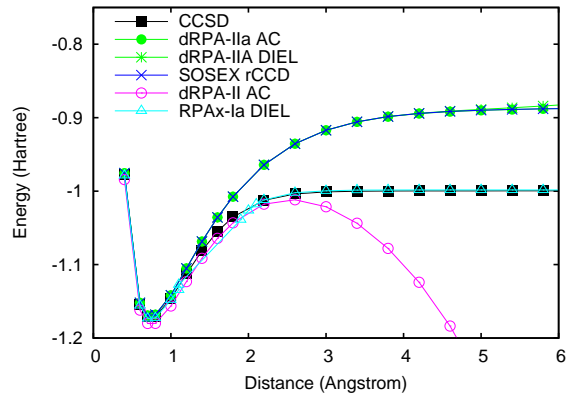}
  \caption{Dissociation curve of H$_2$}\label{fig:H2}
\end{figure}

\begin{turnpage}
\begin{table*}
\caption{\label{tabletot} Differences between total energies computed within different approximations and CCSD(T)/CBS 
reference values (in kcal/mol). Mean error (ME) and
mean absolute error (MAE) of total energies are included in the table. 
The results are presented for the 21 molecules involved in the reactions 
considered in Table~\ref{tablereact} (in kcal/mol). 
The molecules are sorted in increasing order of the absolute values of CCSD(T) total energies.}
\begin{footnotesize}
\begin{tabularx}{\linewidth}{lrrrrrr}
\hline\hline
\makebox[0.15\textwidth][l]{Molecule} &  \makebox[0.15\textwidth][c]{$E_{\text{tot}}^{\text{dRPA-I}}-E_{\text{tot}}^{\text{CCSD(T)}}$}  &     \makebox[0.15\textwidth][c]{$E_{\text{tot}}^{\text{dRPA-IIa}}-E_{\text{tot}}^{\text{CCSD(T)}}$} &   \makebox[0.15\textwidth][c]{$E_{\text{tot}}^{\text{SOSEX}}-E_{\text{tot}}^{\text{CCSD(T)}}$} &   \makebox[0.15\textwidth][c]{$E_{\text{tot}}^{\text{dRPA-II}}-E_{\text{tot}}^{\text{CCSD(T)}}$}  &   \makebox[0.15\textwidth][c]{$E_{\text{tot}}^{\text{RPAx-Ia}}-E_{\text{tot}}^{\text{CCSD(T)}}$} &   \makebox[0.15\textwidth][c]{$E_{\text{tot}}^{\text{CCSD}}-E_{\text{tot}}^{\text{CCSD(T)}}$}   \\
& DIEL\phantom{eccsd} & DIEL\phantom{eccsd} & rCCD\phantom{eccsd} & AC\phantom{deccsd} & DIEL\phantom{eccsd} &  \\
\hline
                            H$_2$ &  -24.67\phantom{eccsd}   &    0.77\phantom{eccsd}  &    0.77\phantom{eccsd}  &    -6.41\phantom{deccsd}  &    -1.92\phantom{eccsd}  &    0.00\phantom{eccsd} \\
                           CH$_4$ &  -92.61\phantom{eccsd}   &    9.34\phantom{eccsd}  &    9.14\phantom{eccsd}  &    -21.57\phantom{deccsd}  &    -1.22\phantom{eccsd}  &    4.60\phantom{eccsd} \\
                           NH$_3$ &  -93.92\phantom{eccsd}   &    12.96\phantom{eccsd}  &    12.70\phantom{eccsd}  &    -18.89\phantom{deccsd}  &    1.56\phantom{eccsd}  &    5.90\phantom{eccsd} \\
                           H$_2$O &  -94.95\phantom{eccsd}   &    15.86\phantom{eccsd}  &    15.58\phantom{eccsd}  &    -15.22\phantom{deccsd}  &    4.05\phantom{eccsd}  &    6.25\phantom{eccsd} \\
                          C$_2$H$_2$ &  -115.23\phantom{eccsd}   &    22.65\phantom{eccsd}  &    22.09\phantom{eccsd}  &    -26.26\phantom{deccsd}  &    6.50\phantom{eccsd}  &    11.82\phantom{eccsd} \\   
                          C$_2$H$_4$ &  -138.49\phantom{eccsd}   &    20.89\phantom{eccsd}  &    20.30\phantom{eccsd}  &    -31.52\phantom{deccsd}  &    2.45\phantom{eccsd}  &    10.80\phantom{eccsd} \\   
                          C$_2$H$_6$ &  -161.23\phantom{eccsd}   &    18.73\phantom{eccsd}  &    18.31\phantom{eccsd}  &    -36.76\phantom{deccsd}  &    -0.26\phantom{eccsd}  &    9.78\phantom{eccsd} \\   
                            CO &  -120.23\phantom{eccsd}   &    26.80\phantom{eccsd}  &    26.14\phantom{eccsd}  &    -23.32\phantom{deccsd}  &    9.69\phantom{eccsd}  &    12.52\phantom{eccsd} \\
                          HCHO &  -142.51\phantom{eccsd}   &    28.09\phantom{eccsd}  &    27.39\phantom{eccsd}  &    -25.65\phantom{deccsd}  &    8.39\phantom{eccsd}  &    12.38\phantom{eccsd} \\
                         CH$_3$OH &  -164.36\phantom{eccsd}   &    25.82\phantom{eccsd}  &    25.31\phantom{eccsd}  &    -30.13\phantom{deccsd}  &    5.21\phantom{eccsd}  &    11.30\phantom{eccsd} \\
                          H$_2$O$_2$ &  -170.33\phantom{eccsd}   &    35.98\phantom{eccsd}  &    35.23\phantom{eccsd}  &    -25.59\phantom{deccsd}  &    12.01\phantom{eccsd}  &    14.42\phantom{eccsd} \\  
                         H$_2$CCO &  -186.86\phantom{eccsd}   &    37.15\phantom{eccsd}  &    36.17\phantom{eccsd}  &    -35.69\phantom{deccsd}  &    11.62\phantom{eccsd}  &    19.21\phantom{eccsd} \\
                         C$_2$H$_4$O &  -210.29\phantom{eccsd}   &    36.61\phantom{eccsd}  &    35.79\phantom{eccsd}  &    -39.63\phantom{deccsd}  &    8.91\phantom{eccsd}  &    17.96\phantom{eccsd} \\   
                         CH$_3$CHO &  -209.91\phantom{eccsd}   &    37.11\phantom{eccsd}  &    36.21\phantom{eccsd}  &    -40.53\phantom{deccsd}  &    9.55\phantom{eccsd}  &    17.82\phantom{eccsd} \\
                        C$_2$H$_5$OH &  -232.47\phantom{eccsd}   &    35.14\phantom{eccsd}  &    34.42\phantom{eccsd}  &    -45.38\phantom{deccsd}  &    6.27\phantom{eccsd}  &    16.74\phantom{eccsd} \\   
                          HNCO &  -188.46\phantom{eccsd}   &    41.28\phantom{eccsd}  &    40.25\phantom{eccsd}  &    -32.07\phantom{deccsd}  &    15.23\phantom{eccsd}  &    21.17\phantom{eccsd} \\
                        HCONH$_2$ &  -210.50\phantom{eccsd}   &    40.25\phantom{eccsd}  &    39.26\phantom{eccsd}  &    -37.01\phantom{deccsd}  &    12.16\phantom{eccsd}  &    19.34\phantom{eccsd} \\
                           CO$_2$ &  -189.37\phantom{eccsd}   &    44.27\phantom{eccsd}  &    43.27\phantom{eccsd}  &    -27.87\phantom{deccsd}  &    18.04\phantom{eccsd}  &    21.19\phantom{eccsd} \\
                         HCOOH &  -211.90\phantom{eccsd}   &    43.75\phantom{eccsd}  &    42.74\phantom{eccsd}  &    -33.33\phantom{deccsd}  &    15.09\phantom{eccsd}  &    19.76\phantom{eccsd} \\
                      NH$_2$CONH$_2$ &  -279.03\phantom{eccsd}   &    52.05\phantom{eccsd}  &    50.85\phantom{eccsd}  &    -48.97\phantom{deccsd}  &    15.47\phantom{eccsd}  &    25.73\phantom{eccsd} \\  
                       HCOOCH$_3$ &  -281.07\phantom{eccsd}   &    53.72\phantom{eccsd}  &    52.47\phantom{eccsd}  &    -48.40\phantom{deccsd}  &    16.35\phantom{eccsd}  &    25.29\phantom{eccsd} \\
\hline
                            ME &  -167.54\phantom{eccsd}   &    30.44\phantom{eccsd}  &    29.73\phantom{eccsd}  &    -30.96\phantom{deccsd}  &    8.34\phantom{eccsd}  &    14.48\phantom{eccsd} \\
                           MAE &  167.54\phantom{eccsd}   &    30.44\phantom{eccsd}  &    29.73\phantom{eccsd}  &    30.96\phantom{deccsd}  &    8.66\phantom{eccsd}  &    14.48\phantom{eccsd} \\
\hline\hline
\end{tabularx}
\end{footnotesize}
\end{table*}
\end{turnpage}

To help the visualization of the results in Table~\ref{tabletot}, we show in Fig.~\ref{totalenergymol} the relative error 
of total energies obtained within different RPA approximations with respect to CCSD(T) results. The dRPA-I DIEL approximation leads to substantially
larger errors and, in order to make the graph more readable, has not been included in the figure. 
The trend of the curves shows that the relative error tends to be approximately constant by increasing the
molecular size.
Since CCSD/CBS is exact for two-electron systems it is not surprising that the corresponding relative error for H$_2$ is zero.
For all the other methods H$_2$ is instead the most problematic case, which is related to the fact that the reference determinant is
constructed from PBE orbitals.
The dRPA-IIa DIEL and SOSEX rCCD approximations give almost identical curves for all the molecules considered here (the difference is at most
1.2 kcal/mol, for NH$_2$CONH$_2$ and HCOOCH$_3$). 
With the exception of H$_2$, the RPAx-Ia relative error is systematically lower than in the CCSD case.
However, as seen in Table~\ref{tablereact}, the accuracy of RPAx-Ia slightly worsens with respect to CCSD in the case of reaction
energies, whose evaluation takes advantage of an error cancellation effect.

In Fig.~\ref{fig:H2}, we show dissociation curves for the singlet state of the  H$_2$ molecule, for which CCSD is exact.
In the literature we can find several studies of H$_2$ dissociation in the context of RPA.
The influence of the reference orbitals has often been studied\cite{Fuchs:05,Dahlen:06,Janesko:09}.
The use of unrestricted Kohn-Sham orbitals is thought to allow for some inclusion of static correlation in the description of the dissociation\cite{Fuchs:05,Dahlen:06}.
Note first that we observed that the Clenshaw-Curtis frequency quadrature is suitable for a certain range of distances only, and that one will need to switch to other quadrature schemes a large distances.
Indeed, for large interatomic distances  H$_2$ is a multireference problem characterized by quasi-degenerate determinants  and dielectric matrix diverges for $\omega \to 0$, hence the currently implemented integration schemes fail.
This is an interesting subject for future work.
The dRPA-II AC diverges at larger distances, which
can also be explained by the fact that the system is dominated here by non-dynamical correlation.
Hence, the coupling-constant integrand is not smooth enough for numerical AC quadratures, as can be inferred from a study of the integrand itself (see for example \cite{Fuchs:05}).
The SOSEX curve shows no bump and is consistent with earlier studies in the literature\cite{Henderson:10,Paier:10,Ren:12,Bates:13}.
The dRPA-IIa AC and dRPA-IIa DIEL, although not theoretically equivalent to SOSEX, are both very close to the SOSEX curve.
Note that the dRPA-IIa DIEL starts to show a deviation from the dRPA-IIa AC results: this is related to the frequency quadrature problem.
The RPAx-Ia-DIEL energy, calculated from a broken-symmetry SCF is very close to the CCSD curve, despite the fact that the fractional deviation between RPAx-Ia-DIEL and  CCSD  
is the worst for H$_2$ among all the  molecules shown on Fig.~\ref{totalenergymol}. Note, however, that the absolute deviation is not large  
in the equilibrium structure
and at dissociation it gets even smaller.

\section{Conclusions, perspectives}\label{conclusions}

The main result of the present work is to show that in contrast to a widely accepted view it is possible to conveniently include exchange effects in the dielectric matrix formulation of RPA correlation energy. Two particular cases have been derived and numerically implemented: the SOSEX-like dRPA-IIa correlation energy and an approximate variant of the RPAx-I method, named RPAx-Ia. 
Our derivation of the dielectric matrix formulation of the RPA proceeds in a "conventional" or "traditional" way, \textit{i.e.}\ from the full adiabatic connection formula at an RPA level. This is in contrast to the work of Eshuis and Furche, who started from the plasmon formula of the dRPA-I energy and derived directly the density fitting expressions. Our approach is strictly equivalent to theirs for dRPA-I but offers in addition a well-defined route to get alternative variants including exchange in a similar form. Furthermore, the connections with the previously studied density matrix formulation are straightforwardly established and therefore our previous results concerning the relationship between the adiabatic connection and rCCD RPA can be simply transferred.

We think that the main interest of the dielectric matrix formalism of RPA in a quantum chemical (more precisely LCAO based) context is mostly conceptual. As demonstrated by the work of Eshuis and Furche, this formalism is well-adapted for density fitting  implementations, which is computationally advantageous, in particular for the dRPA-I case, which is an $O(N^4 \log(N))$ method. The exchange-including variants are expected to show a somewhat less advantageous $O(N^5 \log(N))$ behavior which is comparable to the scaling of DF-MP2~\cite{Werner:03b} or RI-MP2~\cite{Weigend:97,Weigend:02}. The brute-force orbital-based implementations are of $O(N^6)$ scaling.

It is important to notice the new approximations developed within this work may be of significant interest for the condensed matter physics 
community, that mostly
rely on the dielectric matrix formalism to compute RPA correlation energies~\cite{Fuchs:02,Fuchs:05,Harl:08,Lu:09,Nguyen:09,rocca:14}. 

As shown in this work, the relatively poor performance of the dRPA-I is improved to a certain extent by the dRPA-IIa (SOSEX-like) method. 
The approximate RPAx-Ia 
approach is surprisingly good for the systems considered in this paper. In the future more test cases will certainly be necessary to 
fully establish the accuracy of the RPAx-Ia method. Additionally, this methodology is suitable for implementation in plane-wave
codes~\cite{rocca:14,kaoui2016random} and applications to solids will be soon possible.

\section{Aknowledgement}
D.R. acknowledges financial support from Agence Nationale de la Recherche under grant number ANR-15-CE29-0003-01.
The authors thank Andreas Hesselmann for useful discussions and for providing data on the molecular reaction test set.
Computer time was provided
by GENCI-CCRT/CINES  under grant x2015-085106.

\appendix    

\section{Equivalence of the density matrix and logarithmic expressions}
\label{app:dRPADM}

In this appendix we show the equivalence of the density matrix and logarithmic expressions of the dRPA-I correlation energies described in this paper.
The derivation holds for dRPA-IIa, by replacing $\b{K}$ with $\b{B}$.
Let us recall Eq.~(\ref{eq:dRPAIsmall}) as:

\begin{align}
\label{eqapp:dRPAformulas}
 E^{\text{dRPA-I}}_{\text{c}}&=
 \mhalf
 \int_0^1 \!\! d\alpha
 \int_{-\infty}^{\infty}\!\! \frac{d\omega}{2\pi}\,
 \tr
 \biggl\{ \b{\Pi}_{\alpha }(\i\omega)\b{K} -
          \b{\Pi}_0        (\i\omega)\b{K}
 \biggr\}
.\end{align}
Since

\begin{align}
\label{eqapp:Pi0}
\b{{\Pi}}_0(\i\omega) =
 \b{{\Pi}}^+_0(\i\omega) + \b{{\Pi}}^-_0(\i\omega) =
-2 \B{\epsilon}^{1/2} (\B{\epsilon}^2 + \omega^2\b{I})^{-1}
   \B{\epsilon}^{1/2}
,\end{align}
from the {"dimension-reduced" Dyson equation} $ \b{{\Pi}}^{-1}_\alpha(\i\omega)=\b{{\Pi}}^{-1}_0(\i\omega) -\alpha \b{K}$, we have

\begin{align}
\label{eqapp:Piminus1}
\b{{\Pi}}^{-1}_\alpha(\i\omega) = \mhalf
\B{\epsilon}^{-1/2} (\b{M}_\alpha + \omega^2\b{I})
\B{\epsilon}^{-1/2}
,\end{align}
where

\begin{align}
\b{M}_\alpha = \B{\epsilon}^{1/2} (\B{\epsilon} + 2 \alpha \b{K})
           \B{\epsilon}^{1/2}
.\end{align}
The symmetric matrix $\b{M}_\alpha$
can be brought to a diagonal form using its eigenvectors and eigenvalues:

\begin{align}
\b{M}_\alpha \b{Z}_\alpha = \b{Z}_\alpha \b{\Omega}_\alpha^2  ,
\qquad\qquad
\b{Z}_\alpha \b{Z}^\T_\alpha =
\b{Z}^\T_\alpha \b{Z}_\alpha = \b{I} 
.\end{align}
Thus we find:

\begin{align}
\label{eqapp:piminusomega}
\b{{\Pi}}^{-1}_\alpha(\i\omega) & =
\mhalf
\B{\epsilon}^{-1/2}
\b{Z}_\alpha \b{Z}^\T_\alpha
(\b{M}_\alpha + \omega^2\b{1})
\b{Z}_\alpha \b{Z}^\T_\alpha
\B{\epsilon}^{-1/2}
\nonumber \\ & =
\mhalf
\B{\epsilon}^{-1/2}
\b{Z}_\alpha
 (\b{\Omega}^2_\alpha + \omega^2\b{I})
 \b{Z}^\T_\alpha
\B{\epsilon}^{-1/2}
,\end{align}
and furthermore:

\begin{align}
 \b{{\Pi}}_\alpha(\i\omega) =
 -2
 \B{\epsilon}^{1/2}
\b{Z}_\alpha
 (\b{\Omega}^2_\alpha + \omega^2\b{I})^{-1}
 \b{Z}^\T_\alpha
\B{\epsilon}^{1/2}
.\end{align}

Noting that

\begin{align}
\int_{-\infty}^{\infty}\!\!\frac{d\omega}{2\pi}\,
\left(\Omega_{\alpha,ia}^2 +\omega^2\right)^{-1} =
\half \Omega_{\alpha,ia}^{-1}
,\end{align}
where $ia$ labels the eigenvalues of $\b{M}_\alpha$, \textit{i.e.} the diagonal elements of $\b{\Omega}$, we get

\begin{align}
\int_{-\infty}^{\infty}\!\!\frac{d\omega}{2\pi}\,
\b{{\Pi}}_\alpha(\i\omega) & =
-\B{\epsilon}^{1/2}
\b{Z}_\alpha
\b{\Omega}_{\alpha}^{-1}
\b{Z}^\T_\alpha
\B{\epsilon}^{1/2} =
-
\B{\epsilon}^{1/2}
\b{M}_{\alpha}^{-1/2}
\B{\epsilon}^{1/2}
.\end{align}
Furthermore, since $\b{M}_0=\B{\epsilon}^2$, we have

\begin{align}
\int_{-\infty}^{\infty}\!\!\frac{d\omega}{2\pi}\,
\b{{\Pi}}_0(\i\omega) =
-
\B{\epsilon}^{1/2}
\B{\epsilon}^{-1}
\B{\epsilon}^{1/2}= -\b{I}
,\end{align}
which finally leads to

\begin{align}
\b{E}^\text{dRPA-I}= \half \int_0^1 d\alpha\,
\b{P}_{c,\alpha}\b{K}
,\end{align}
with

\begin{align}
\b{P}_{c,\alpha}= \B{\epsilon}^{1/2}
\b{M}_{\alpha}^{-1/2}
\B{\epsilon}^{1/2} - \b{I}
.\end{align}

This derivation demonstrates the equivalence of the density matrix and dielectric matrix formulations of the dRPA-I (and \textit{mutatis mutandis}, of the dRPA-IIa) energy expressions.

\section{Plasmon formula}\label{app:plasmon}

Starting from Eq.~(\ref{eq:dRPAIlog}), we re-write the argument of the logarithm by using the {"dimension-reduced"} Dyson equation in the following form:

\begin{align}
  \b{I}-\b{{\Pi}}_0(\i\omega )\,\b{K} = \b{{\Pi}}_0(\i\omega ) \b{{\Pi}}_1^{-1}(\i\omega)
.\end{align}
Combining Eqs.(\ref{eqapp:Pi0}) and (\ref{eqapp:piminusomega}), one obtains

\begin{align}
\label{eqapp:Pi0PiMinus1}
\b{{\Pi}}_0(\i\omega) \b{{\Pi}}^{-1}_1(\i\omega) & =
 \B{\epsilon}^{1/2} (\B{\epsilon}^2 + \omega^2\b{I})^{-1}
 \b{Z}_1(\B{\Omega}_1^2+\omega^2\b{I})\b{Z}_1^\T
 \B{\epsilon}^{-1/2}
.\end{align}
Employing
$\tr\left\{\log(\b{X})\right\}
=\log\left\{\det(\b{X})\right\}$
and properties of the determinant leads to:

\begin{align}
&  \tr  \left\{
   \log \left(
     \b{{\Pi}}_0(\i\omega) \b{{\Pi}}_1^{-1}(\i\omega)
   \right)
   \right\}
\nonumber\\&\quad
=
 \log \left\{
 \det \left(
   (\B{\epsilon}^2 + \omega^2\b{I})^{-1}
   (\b{\Omega}_1^2 + \omega^2\b{I})
 \right)
 \right\}
\nonumber \\&\quad
  =
 \sum_{ia}
 \log\left(
 \frac{\Omega_{1,ia}^2+\omega^2}
      {\epsilon^2_{ia}+\omega^2}\right) =
 \sum_{ia}
 \log
 \left(1 +\frac{\Omega_{1,ia}^2-\epsilon^2_{ia}}
                {\epsilon^2_{ia}+\omega^2}
 \right)
.\end{align}

For the evaluation of the trace of $\b{{\Pi}}_0(\i\omega )\b{K}$ we use the definition of $\b{M}_1$ to write $\b{K}= \tfrac{1}{2}\B{\epsilon}^{-1/2} (\b{M}_1 - \B{\epsilon}^2)\B{\epsilon}^{-1/2}$ and we obtain

\begin{align}
  \tr\left\{
  \b{{\Pi}}_0(\i\omega )\,\b{K}\right\} & =
  -\tr
   \left\{
    (\B{\epsilon}^2 + \omega^2 \b{I})^{-1}
   \left(\b{M}_1 -
   \B{\epsilon}^2\right)
   \right\}
.\end{align}

Summarizing, one finally obtains:

\begin{align}
\label{appeq:dRPAlogint}
   E_{\text{c}}^{\text{dRPA-I}}=
   \half
   \int_{-\infty}^{\infty}\!\! \frac{d\omega}{2\pi}\,
   \sum_{ia}
  \left\{
   \log
   \left( 1 +\frac{\Omega_{ia}^2 -\epsilon_{ia}^2}
                 {\epsilon_{ia}^2 + \omega^2}
   \right) -
   \frac{M_{1,ia,ia} -\epsilon_{ia}^2}
        {\epsilon_{ia}^2 + \omega^2}
   \right\}
,\end{align}
which, after integration (see Sec.~1 of the Supporting Information)
, becomes

\begin{align}
\label{appeq:dRPAplasmon}
   E_{\text{c}}^{\text{dRPA-I}}=
   \half \biggl\{\sum_{ia}^{N_\text{exc}}\Omega_{ia} -
   \sum_{ia}^{N_\text{exc}} (\epsilon_{ia} + K_{ia,ia})\biggr\} 
.\end{align}

\section{Density-based implementation of dRPA}
\label{app:DF}

The dRPA correlation energy can be brought to a computationally more efficient form
by using density fitting (sometimes called resolution-of-identity)
or Cholesky decomposition methods
applied to the two-electron integrals. In these techniques the $(N_\text{exc}\times N_\text{exc})$ two-electron integral matrix is decomposed  as
(see also Sec.~3 of the Supporting Information)

\begin{align}
\label{eq:DFdecomp}
\b{K}_{ia,jb} = \left[ \b{M} \b{M}^\T \right ]_{ia,jb}=  2\sum_G L_{ia,G} L_{G,jb}
,\end{align}
where $\b{M}$ is a $(N_\text{exc}\times N_\text{aux})$ rectangular matrix, and $N_\text{aux}$ is significantly smaller than $N_\text{exc}$ (but larger than $N_\text{occ}+N_\text{virt}$). As shown in Sec.~3 of the Supporting Information
, the correlation energy Eq.~(14) of our paper 
becomes

\begin{align}\label{eq:dRPAIlogDF}
   E_{\text{c}}^{\text{dRPA-I}}=
   \int_0^{\infty}\!\! \frac{d\omega}{2\pi}\,
   \tr
   \left\{
   \log \left(\b{1}-\b{C}(\i\omega)\right) +
   \b{C}(\i\omega)
   \right\}
,\end{align}
where $\b{C}(\i\omega) = \b{M}^\T\,\b{{\Pi}}_0(\i\omega )\,\b{M}$ is a $(N_\text{aux}\times N_\text{aux})$ matrix
and $\b{1}$ is here the unit matrix of dimension $N_\text{aux}$.
Remembering that $\b{{\Pi}}_0(\i\omega ) = -2 \B{\epsilon}^{1/2}(\B{\epsilon}^2 + \omega^2\b{I})^{-1} \B{\epsilon}^{1/2}$ it  is easy to see that Eq.~(\ref{eq:dRPAIlogDF}) is the same expression as the one obtained in an elegant but relatively involved manner by Eshuis  \textit{et al.}~\cite{Eshuis:10}, starting from the plasmon expression of the correlation energy and the integral representation of the square root of a matrix.

The density fitting technique can be generalized to the second order screened exchange correction in the dRPA-IIa correlation energy, leading to the following expression (see Sec.~3 of the Supporting Information)

\begin{align}
\label{eq:dRPAIIDF}
E_{\text{c}}^{\text{dRPA-IIa}} 
&=
  \int_0^{\infty}\!\!\frac{d\omega}{2\pi}\,
 \tr\biggl\{\bigl(\log\bigl(\b{1}-\b{C}(\i\omega)\bigr)+\b{C}(\i\omega)\bigr) 
\times\nonumber\\&\qquad\qquad\qquad
 \bigl(\b{1}
 - 
   \b{C}^{-1}(\i\omega) \b{Y}(\i\omega)\b{C}^{-1}(\i\omega)\bigr)
 \biggr\}
,\end{align}
and the corresponding density fitting Casimir-Polder transform MP2  energy is obtained as the second-order contribution  $(n=2)$ in Eq.~(\ref{eq:dRPAIIDF}):

\begin{align}
E_{\text{c}}^{\text{MP2}} & =
  \mhalf
  \int_0^{\infty}\!\!\frac{d\omega}{2\pi}\,
 \tr\biggl\{\b{C}^2(\i\omega)
 -
   \b{Y}(\i\omega)
 \biggr\}
.\end{align}

As shown in the Sec.~3 of the Supporting Information
,
the matrices $\b{C}(\i\omega)$
and $\b{Y}(\i\omega)$
are defined in terms of the
matrices
appearing in the decomposition of the two-electron integrals seen in Eq.~(\ref{eq:DFdecomp}) and of the orbital energy differences:

\begin{align}
\b{C}(\i\omega)=2 \sum_i \b{X}^{ii}(\i\omega)
,\end{align}
and

\begin{align}
\b{Y}(\i\omega)= 2\sum_{ij}
  \b{X}^{ij}(\i\omega)\b{X}^{ij}(\i\omega)
,\end{align}
with

\begin{align}
[\b{X}^{ij}(\i\omega)]_{PQ}=
   \sum_{a}
   L_{P,ia}
   \frac{
-2 \epsilon_{ia}
 }{
 \epsilon_{ia}^2+\omega^2}
 L^{~}_{ja,Q}
.\end{align}

In the density fitting case the working equations are similar to those in the orbital-based implementation.
They can be obtained after diagonalization of the $\b{C}(\i\omega)$ matrix
and take advantage of the fact that the dimensions $(N_\text{aux}\times N_\text{aux})$ of the matrix to be diagonalized is considerably smaller than $N_\text{exc}\times N_\text{exc}$.

%
%

\bibliography{MusRocJanAng-2015}
\end{document}